\documentclass{aa}

\usepackage{txfonts}
\usepackage{natbib}
\usepackage{graphicx}
\usepackage{aalongtable}
\usepackage{calc}

\bibpunct{(}{)}{;}{a}{}{,} 

\def\etal{et al.\/}

\def\muere{\mbox{$\langle\mu\rangle_{\rm e} - \log(R_{\rm e})$}}

\begin{document}


\title{WINGS: a WIde-field Nearby Galaxy-cluster Survey. I:\\ Optical imaging}

\author{
G. Fasano$^1$ \and C. Marmo$^{2,3}$ \and J. Varela$^1$ \and M. D'Onofrio$^4$ \and B.M. Poggianti$^1$ \and M. Moles$^5$ \and \\
E. Pignatelli$^1$ \and D. Bettoni$^1$ \and P. Kj{\ae}rgaard$^6$ \and L. Rizzi$^7$ \and W.J. 
Couch$^8$ \and A. Dressler$^9$
}

\institute{
$^1$ INAF -- Padova Astronomical Observatory, Vicolo Osservatorio 5, 35122 Padova, Italy\\
$^2$ CEA/DSM/DAPNIA, Service d'Astrophysique, Gif-sur-Yvette, France\\
$^3$ Institut d'Astrophysique de Paris, 98 bis bd Arago, 75014 Paris, France\\
$^4$ Astronomy Department, University of Padova, Vicolo Osservatorio 2, 35122 Padova, Italy\\
$^5$ Instituto de Astrof\'{\i}sica de Andaluc\'{\i}a (C.S.I.C.) Apartado 3004, 18080 Granada, Spain\\
$^6$ Copenhagen University Observatory. The Niels Bohr Institute for Astronomy Physics and
            Geophysics, Juliane Maries Vej 30, 2100 Copenhagen, Denmark\\
$^7$ Institute for Astronomy, University of Hawaii, 2680 Woodlawn Drive, Honolulu, Hi 96822, USA\\
$^8$ School of Physics, University of New South Wales, Sydney 2052, Australia\\
$^9$ Observatories of the Carnegie Institution of Washington, Pasadena, CA 91101, USA\\
}  

\offprints{Giovanni Fasano,\\ \email{fasano@pd.astro.it}}
\date{ Received 13~July~2005 / Accepted 5~September~2005 }

\abstract{
This is the first paper of a series that will present data and
scientific results from the WINGS project, a wide-field,
multiwavelength imaging and spectroscopic survey of galaxies in 77
nearby clusters. The sample was extracted from the ROSAT catalogs of
X-Ray emitting clusters, with constraints on the redshift
($0.04< z<0.07$) and distance from the galactic plane
(${\vert}b{\vert}\geq$20~deg).
 
The global goal of the WINGS project is the systematic study of the
local cosmic variance of the cluster population and of the properties
of cluster galaxies as a function of cluster properties and local
environment. This data collection will allow the definition of a local,
'zero-point' reference against which to gauge the cosmic evolution when 
compared to more distant clusters.

The core of the project consists of wide-field optical imaging of
the selected clusters in the $B$ and $V$ bands. We have also completed
a multi-fiber, medium-resolution spectroscopic survey for 51 of the
clusters in the master sample. The imaging and spectroscopy data were 
collected using, respectively, the
WFC@INT and WYFFOS@WHT in the northern hemisphere, and the WFI@MPG and
2dF@AAT in the southern hemisphere. In addition, a NIR ($J$,$K$) survey of $\sim$ 50
clusters and an $H_{\alpha}+U$ survey of some 10 clusters are presently ongoing
with the WFCAM@UKIRT and WFC@INT, respectively, while a very-wide-field
optical survey has also been programmed with OmegaCam@VST.

In this paper we briefly outline the global objectives and the main
characteristics of the WINGS project. Moreover, the observing strategy
and the data reduction of the optical imaging survey (WINGS-OPT) are
presented. We have achieved a photometric accuracy of $\sim$0.025~mag,
reaching completeness to V$\sim$23.5. Field size and resolution (FWHM)
span the absolute intervals (1.6-2.7)~Mpc and (0.7-1.7)~kpc,
respectively, depending on the redshift and on the seeing. This allows
the planned studies to obtain a valuable description of the local
properties of clusters and galaxies in clusters.

}

\titlerunning{WINGS I: Optical imaging of clusters}
\maketitle

\section{Introduction}
\label{sec:Intro}

Galaxies of different morphology are not evenly distributed. 
It is now more than 70 years since \citet{hubb} first
noticed that (in the local universe) spiral
galaxies are abundant in the field while S0 and elliptical galaxies
dominate in denser regions. Gravitational interaction apparently affects the
global properties of the galaxies even in low density environments, and
even such field galaxies show significant differences with respect to
truly isolated systems that have been free of interaction for a long period
of time \citep{vare1}.

Clusters of galaxies are dense peaks in the galaxy distribution and
therefore appropriate sites to look for changes in the properties of
the galaxies. They can be therefore used to trace the evolution of the
systems themselves as well as that of the galaxies in them. Such a
systematic analysis certainly needs a fair knowledge of
the properties of local clusters of galaxies and their content (the
end point of the evolution), extensive enough to cope not only with the
average properties but also with their physical variance. This is
unfortunately still lacking.  As a matter of fact, while a large
amount of high quality data for distant clusters is continuously being
gathered from both HST observations and large ground-based telescopes,
our present knowledge of the systematic properties of galaxies in
nearby clusters, remains surprisingly limited, with Virgo, Coma and
Fornax as the main references.

In the range 0.4$\leq z \leq$0.5, exploiting the high spatial
resolution achieved with the Hubble Space Telescope (HST), \citet{dres2} 
and \citet{smai} found that spirals are a factor
of 2-3 more abundant and S0 galaxies are proportionally less abundant
than in nearby clusters, while the fraction of ellipticals is already
as large or larger. This implies significant morphological
transformations occurring rather recently. Similarly, using 
excellent-seeing, ground based imaging with the NOT telescope (La
Palma), \citet{fasa1} completed the picture in the range
0.1 $\leq z \leq$ 0.25, showing that the S0 population smoothly
grows from $z\sim 0.5$ to $z\sim 0$, at the expense of the population
of spiral galaxies. They also highlighted the role that the {\em
cluster type} plays in determining the relative occurrence of S0 and
elliptical galaxies at a given redshift: clusters at $z\sim 0.1-0.2$
have a low (high) S0/E ratio if they display (lack) a strong
concentration of elliptical galaxies towards the cluster centre. This
dichotomy seems to support Oemler's (1974) suggestion that
elliptical-rich and S0-rich clusters are not two evolutionary stages
in cluster evolution, but intrinsically different types of clusters in
which the abundance of ellipticals was established at redshifts much
greater than $0.5$.

That trend is supported by the morphological studies at $z > 0.5$,
that find an even lower fraction of early-type galaxies (Es+S0s), thus
indicating that this fraction keeps decreasing up to $z\sim 1$
\citep[ Simard \etal\, private communication]{dokk,lubi2}. The
most recent works, based on the Advanced Camera for Surveys,
demonstrate that it is the decreasing proportion of S0 galaxies that
drives this decline also at $z \sim 0.8 - 1$ 
\citep[ Desai \etal\, in preparation]{post}. 
This change of the morphological mix in clusters
is expressed in the evolution of the morphology-density
relation with $z$ \citep{dres2,post}. 

The work on intermediate-redshift clusters observed by HST has been
complemented with ground-based spectroscopic surveys that have led to a
detailed comparison of the spectral and morphological properties 
\citep{dres3,pogg,couc1,couc2,couc3,fish,lubi1,balo1,balo2}.
These studies have shown that the spiral population includes
most of the star-forming galaxies, a large number of post-starburst
galaxies and a sizeable fraction of the red, passive galaxies; in
contrast, the stellar populations of (the few) S0 galaxies appear to
be as old and passively evolving as those in the ellipticals. These
observations are consistent with the post-starburst and star-forming
galaxies being recently infallen field spirals whose star formation is
truncated upon entering the cluster and that will evolve into S0's at
a later time.

At variance with intermediate redshift clusters, for which recent,
high-quality photometric data are available, the morphological
reference for local clusters is still the historical database of
\citet{dres1}, based on photographic plates, giving the positions,
the estimated magnitudes (down to $V\sim 16$) and the visual
morphological classification for galaxies in 55 clusters in the range
0.011 $\leq z\leq$ 0.066. This awkward situation can be easily
understood since only with the new large format (wide-field) CCD mosaic
cameras a significant number of low redshift clusters could be
reasonably well mapped.

Our goal has been to help fill this information gap. Accordingly,
we began in 1999 a program to secure a large database for a local
sample of clusters, to study the cosmic variance of the cluster
properties and their populations in a systematic way. The result would
be a reference 'zero-point' for comparison with studies at higher $z$ and
for evolutionary studies. To that end we have collected wide-field
photometric and spectroscopic data for an X-ray selected sample of 77
clusters at low redshift, spanning a wide range in X-ray and optical
properties. The observational requirements have been set to ensure an
adequate data quality, both for imaging and spectroscopy, in order to
obtain detailed and reliable morphological classifications and
estimates of stellar population ages, metallicities and star formation
histories.

Similar projects were, in the meantime, also begun, either for smaller
samples \citep{pimb,chri}, or
with more limited goals \citep{smit,nela}. On the
spectroscopic side, the ESO Nearby Abell Cluster Survey \citep[ ENACS]{katg1,bivi1}
collected redshifts for
galaxies in 107 clusters, of which 67 with at least 20 spectroscopic
members. This dataset yielded information on cluster velocity
dispersions, kinematics and spatial distributions of different types
of galaxies, that motivated detailed analysis of cluster properties
\citep{katg2,bivi1,bivi2,mazu}. Samples of low-redshift clusters have been also identified
based on the redshifts obtained by two recent large spectroscopic
surveys, 2dF and Sloan \citep{depr,nich,voge}, 
the former having no corresponding CCD imaging database. Results based on
these surveys have highlighted the strong correlation between star
formation properties in galaxies and local galaxy density, and that
such a correlation exists both inside and outside of clusters 
\citep{lewi,gome,kauf,balo3}. Unfortunately, given the typical spatial resolution of the
imaging data and the magnitude limit of the SDSS, it is not immediately
possible to make a detailed comparison with the existing high redshift
morphological and spectroscopic studies.

This paper is the first of a series presenting the results
of this project, that we have called WINGS for {\sl WIde-field Nearby
Galaxy-cluster Survey}. The goal of the present paper is to outline
the objectives and the main characteristics of the WINGS program
(Section~\ref{sec:WINGS}) and to describe in detail the optical imaging
observations. The selection of the cluster sample is presented in
Section~\ref{sec:CluSam}, while Section~\ref{sec:Obs}
is devoted to the description of the observations and the procedures
for the reduction of the optical wide-field survey (WINGS-OPT). The
data quality of optical imaging is analysed in
Section~\ref{sec:Quality}. Finally, a brief summary of the future plans
concerning the whole WINGS project is given in
Section~\ref{sec:FutPlans}.

In this and in the following papers of the series we assume the now
standard metric with $H_0$=70, $\Omega_m$=0.3 and
$\Omega_\Lambda$=0.7.


\section{The WINGS project}
\label{sec:WINGS}

\begin{figure}
\vskip -0.5truecm
\begin{center}
\hskip -1truecm
\includegraphics[width=9cm]{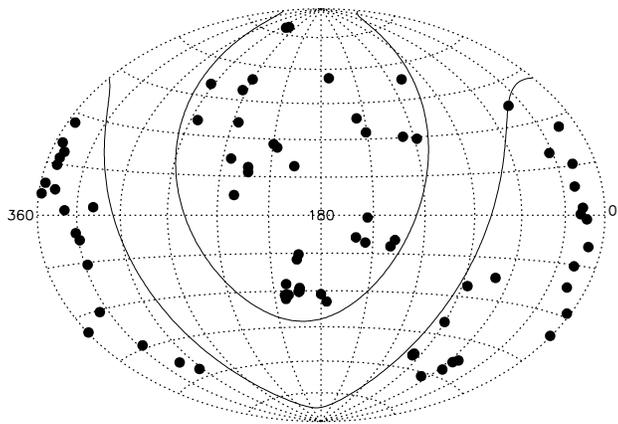}
\vskip -1.5truecm
\caption{All-Sky Aitoff map of the cluster sample (equatorial coordinates).
Lines delimiting the region $\vert b\vert\leq$20 are drawn.}
\label{fig:aitoff}
\end{center}
\end{figure}

The principal goal of the WINGS project is to elaborate a
statistically meaningful, high quality database of the properties of
nearby clusters of galaxies and of the galaxies that populate
them. Hopefully, this will serve to improve our knowledge of
clusters and cluster galaxies in the local universe and
will provide the reference to gauge the changes with redshift over
their physical variance at a given $z$.

In broad terms, the goals of the project are to characterize the
global properties of clusters taken as systems, and those of their
member galaxies. Among the former, besides the already existing 
data on the X-ray luminosity, we include their total luminosity
and size, the velocity dispersion, the presence of substructures and
the cluster scaling relations \citep{marm2}. This will allow us
to explore the existence of well defined relations among
structural parameters and characterize the actual range of those
properties.

Regarding the member galaxies, our primary goals are to analyze the
variance of the morphological fractions (E/S0/S/Irr), their
distribution in the clusters and the morphology-density relation. The
analysis of the colors and the spectral information will provide the
data necessary to retrace the star formation history
of galaxies in nearby clusters.

The WINGS project was designed to cover all these topics. Originally it
was planned as a wide-field optical ($B$,$V$) imaging survey. This is the core of
the project, hereafter called WINGS-OPT. The strategy for imaging
and for the resulting data reduction are the main subject of the
present article.

In addition, other surveys were designed and carried out to complement the
characterization of the cluster galaxies. The already completed 
WINGS-SPE survey consists of
multi-fiber spectroscopy of galaxies in 51 clusters from the master
WINGS sample, obtained with the WYFFOS@WHT and the 2dF@AAT
spectrographs over the same area covered by the optical imaging
($34'\times34'$). The spectra cover the range 3800-7000 \AA $\,$
(WYFFOS) and 3600-8000 \AA $\,$ (2dF), with dispersions of 3\AA\ and 9\AA,
respectively, for the galaxies with $V < 20$ (between 100 and 300 per
cluster). This limit is 1.5 and 2.0 mag deeper than the 2dF
and Sloan surveys, respectively.

Three more follow-up surveys of clusters in the WINGS sample are
presently ongoing. The first one is a NIR (WINGS-NIR: $J$ and $K$-bands)
imaging survey, with the
new Wide-Field Camera at the 3.8\,m UKIRT telescope. This will obtain
data for $\sim 50$ clusters, useful at providing an
estimate of the stellar mass of galaxies, as well as constraining the
spectral energy distribution of galaxies in these fields. The other ones
are $H_{\alpha}$ and U-broad-band surveys (WINGS-HAL and WINGS-UV, 
respectively), with the WFC@INT camera and purpose-defined narrow-band 
filters (for the WINGS-HAL survey), to image $\sim$1 square degree of
10 WINGS clusters. Finally, a very-wide-field ($\sim$1 square degree)
optical survey (WINGS-VWF), with the ESO-VST telescope, equipped with
OmegaCam, has been programmed for the near future.

In combination, these data will constitute a multiwavelength photometric
and spectroscopic dataset which will allow detailed studies of the
properties of nearby Clusters of Galaxies, and cope with their
variance, necessary to identify the cosmic evolution when compared
with those of higher redshift systems.

We present here the observations, data reduction and analysis of data
quality from WINGS-OPT. For all galaxies down to the limit of
detectability we have extracted the position, size, concentration,
average flattening and orientation, as well as the integrated and
aperture photometry in the two observed bands, $B$, $V$. For a
subsample of large galaxies we have also obtained detailed surface
photometry (luminosity and geometrical profiles) and global structural
parameters (total magnitudes, effective radii, ellipticity and
S\'ersic index) using our automatic surface photometry tool GASPHOT
\citep{pign}. Finally, morphological type
estimates of the same subsample of large galaxies, compared and
calibrated with visual classifications, were automatically obtained
with the purpose-written tool MORPHOT \citep{fasa3}.

The catalogues and the statistical analyses of galaxies and 
cluster properties will be
presented in subsequent papers of this series. To maximize the
scientific outcome of the data, the whole WINGS dataset and products,
including photometry, surface photometry, morphological and
spectroscopic catalogs, will become publicy available as the
corresponding papers of this series are published.

\section{The cluster sample}
\label{sec:CluSam}

\begin{figure}
    \includegraphics[width=9cm]{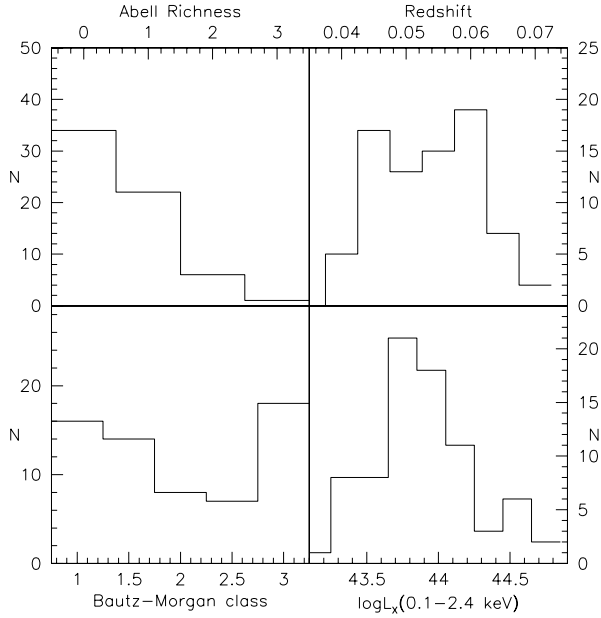}
  \caption{Distribution of some cluster properties in the WINGS sample}
  \label{fig:SampDis}
\end{figure}

To investigate in a systematic way the correlations
between cluster properties and cluster galaxy populations, a
well-defined, large cluster sample is required, with available X-ray
data and covering a wide range in optical and X-ray properties.

WINGS clusters have been selected from three X-ray flux limited samples
compiled from ROSAT All-Sky Survey data: the ROSAT Brightest Cluster
Sample \citep[ BCS]{ebel2}, and its extension \citep[ eBCS]{ebel3}
in the Northern hemisphere and the X-Ray-Brightest Abell-type
Cluster sample \citep[ XBACs]{ebel1} in the Southern
hemisphere. These catalogs are uncontaminated by non-cluster X-ray
sources (AGNs or foreground stars).  The BCS is 90\% complete for
fluxes higher than $4.4 \times 10^{-12} \, \rm erg \, cm^{-2} \,
s^{-1}$ in the 0.1-2.4 keV band. The eBCS extends the BCS down to $2.8
\times 10^{-12} \, \rm erg \, cm^{-2} \, s^{-1}$ with 75\%
completeness. Finally, the XBACs is an essentially complete sample of
Abell clusters with fluxes above $5.0 \times 10^{-12} \, \rm erg \,
cm^{-2} \, s^{-1}$.\footnote{Note that our sample largely overlaps
with the one studied by \citet{smit}.}

The original WINGS sample comprises all clusters from BCS, eBCS and
XBACs with a high Galactic latitude ($|b|\ge 20$ deg) in the redshift
range $0.04< z< 0.07$. The redshift cut and the Galactic latitude are
thus the only selection criteria applied to the X-ray samples.  The
redshift range has been chosen to guarantee both a large area coverage
(the side of our field is $34' \ge 1.6 \, \rm Mpc$) and sufficient
spatial resolution ($1'' \le 1.3 \, \rm kpc$) for all clusters.

After having removed the cluster A3391, because of the
presence of strong non-uniform illumination in the CCD frames, the
final WINGS sample includes 77 clusters (41 in the Southern Hemisphere
and 36 in the Northern Hemisphere, see Figure~1), of which 18 are 
in common with Dressler's (1980) sample. This partial overlap is useful 
for comparing the two datasets and the morphological classifications.
Table~\ref{tab:Sample} (Online Material) lists the cluster name, coordinates of the
adopted center, redshift, Abell richness, Bautz-Morgan type, X-Ray
luminosity from Ebeling \etal\ (1996,1998,2000; converted to our cosmology) 
in units of $10^{44}$~erg~$s^{-1}$ and color excess E(B-V). 

The WINGS clusters span a wide range in X-ray luminosities ($\rm log
\, L_X[0.1-2.4 \, keV]= 43.2-44.7$), corresponding to $\sim 5 \times 10^{14}
\rm \, to \, > 10^{15}$ gravitational solar masses
\citep{reip}, as well as in optical properties such as Abell richness
and Bautz-Morgan type (see Figure~\ref{fig:SampDis}).

\begin{table*}
\caption{The WINGS-OPT observing runs}
\label{tab:ObsRun}
\begin{tabular}{cccccc}
\hline\hline
\multicolumn{6}{c}{WFC INT-2.5m (North)} \\
\hline
Run number & PATT/CAT~REF. & Starting Date & Alloc.Time & $B$ Filt.ID & $V$ Filt.ID \\
\hline
1 & C3   & Aug. 28, 2000 & 1 night  & Harris~(191) & Harris~(192) \\
2 & ITP3 & Apr. 25, 2001 & 5 nights & Kitt~Peak~(210) & Harris~(192) \\
4 & ITP3 & Sep. 15, 2001 & 3 nights & Harris~(191) & Harris~(192) \\
\hline
\hline
\multicolumn{6}{c}{WFI MPG/ESO-2.2m (South)} \\
\hline
Run number & Proposal ID & Starting Date & Alloc.Time & $B$ Filt.ID & $V$ Filt.ID \\
\hline
3 & 67.A-0030 & Aug. 15, 2001 & 2 nights & ESO99~(842) & ESO89~(843) \\
5 & 68.A-0139 & Feb. 14, 2002 & 30 hrs   & ESO99~(842) & ESO89~(843) \\
6 & 69.A-0119 & Apr. 01, 2002 & 18 hrs   & ESOnewB~(878) & ESO89~(843) \\
\hline
\end{tabular}
\end{table*}

\section{WINGS-OPT survey}
\label{sec:Obs}

\subsection{Survey requirements}
\label{sec:SurvReq}

\begin{figure}
    \includegraphics[width=9cm]{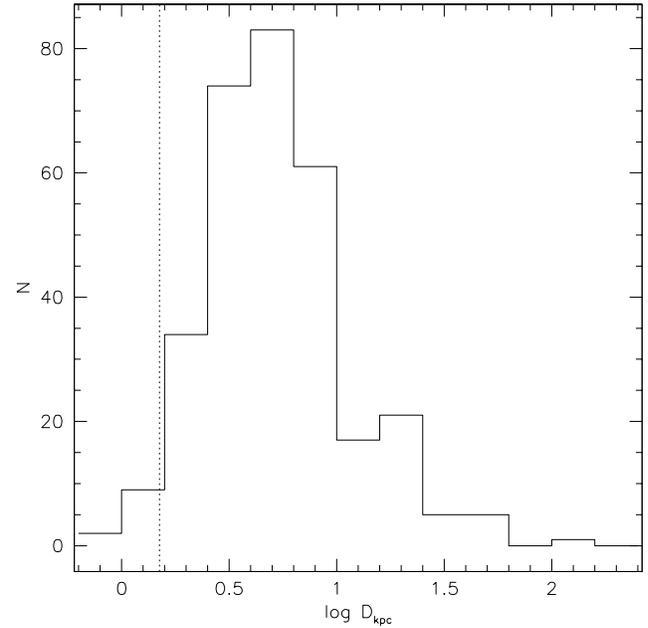}
  \caption{Distribution of effective diameters (in kpc) for early-type galaxies
in the nearby clusters studied by \citet{fasa2}. The dotted line corresponds
to 1.5~kpc in our cosmological framework.}
  \label{fig:Dhist}
\end{figure}

Among the attributes of any photometric galaxy survey, the most
important ones concern the spatial resolution and the photometric
depth. Concerning the former, Figure~\ref{fig:Dhist} shows the
distribution of effective diameters (in kpc) for early-type galaxies
in the nearby clusters studied by \citet{fasa2}. Since a good
galaxy profile restoration is usually possible down to effective
diameters of the order of the Full Width Half Maximum (FWHM) of the
point spread function (see Figure~4 in \citealt{fasa2}), we chose
as the WINGS-OPT imaging requirement that the FWHM not exceed
$\sim$1.5~kpc in our cosmological framework (the dotted line in
Figure~\ref{fig:Dhist}).

Concerning the photometric depth, our interest is twofold: First, we
want the WINGS-OPT survey to be able to sample the luminosity function
of clusters down to the dwarf galaxies ($M_V\sim$~-14). Second, we
require that the depth is sufficient to allow a reliable surface
photometry (S/N ratio $\approx$4.5 per square arcseconds) down to a
surface brightness of $\mu_V\sim$25~mag~arcsec$^{-2}$.
Section~\ref{sec:Quality} illustrates to what extent the above
mentioned requisites have been fulfilled by the WINGS-OPT
observations.

\subsection{Observations}
\label{sec:Obser}

The observations of the WINGS-OPT survey have been taken in dark time
with the Wide~Field~Camera (WFC) mounted at the corrected f/3.29 prime
focus of the INT-2.5m telescope in La~Palma (Canary Islands, Spain)
and with the Wide~Field~Imager (WFI) mounted at the f/8 Cassegrain
focus of the MPG/ESO-2.2m telescope in La~Silla (Chile) for the
northern and southern clusters, respectively. The northern campaign
consisted of three runs, totalling 9 nights, during which 46 clusters
were observed.  The southern campaign has produced data for 35 clusters
during three observing runs (the last two in service mode), for a
total of 2 nights in observer mode, plus about 48 hours of science
exposures in service mode. Tables~\ref{tab:ObsRun} and
~\ref{tab:WFCWFI} list the observing runs of the WINGS-OPT survey and
the main instrumental characteristics of the wide-field cameras,
respectively.

\begin{table}
\caption{Technical features of the wide-field cameras used by the WINGS-OPT survey}
\label{tab:WFCWFI}
\begin {tabular}{lcc}
\hline\hline
Feature & WFC@INT & WFI@MPG \\
\hline
 Field of view & $34'\times34'$ & $34'\times33'$\\
 Pixel scale & 0.33$^{\prime\prime}$/pixel & 0.238$^{\prime\prime}$/pixel \\
 Detector & $4\times 2k\times 4k$ & $8\times 2k\times 4k$ \\
 Filling factor & $93.6\%$ & $95.9\%$ \\
 Read-out noise & $6.2~e^-$/pix & $4.5~e^-$/pix \\
 (Inverse) gain & $2.8~e^-/ADU$ & $2.0~e^-/ADU$ \\
 Full-well capacity & $\sim180,000~e^-$ & $>200,000~e^-$ \\
 Telescope aperture & $2.54~m$ & $2.20~m$ \\
 Telescope focus & Prime~focus & Cassegrain \\
\hline
\end{tabular}
\end{table}

\begin{figure}
    \includegraphics[width=9cm]{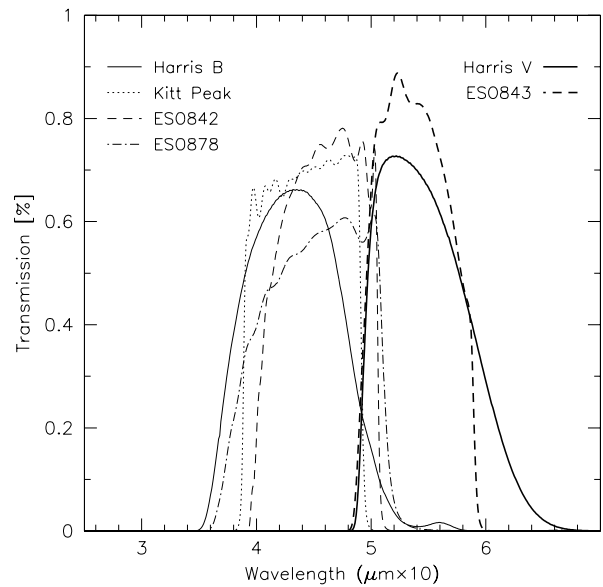}
  \caption{Transmission curves of the filters used in the WINGS-OPT survey}
  \label{fig:Filt}
\end{figure}

We decided to take images in the $V$ and $B$ bands. The V filter allows
us to compare our results with previous studies of nearby clusters, as
well as with WFPC2/ACS@HST ($F814W$) studies of clusters at
$z\sim$0.5. The B filter is needed in order to get colors of
galaxies and especially useful because it is the rest-frame
equivalent to the imaging of clusters at $z\ge$0.5 done using HST~+~ACS.
Table~\ref{tab:ObsRun} reports the identifications of the broad band
$B$ and $V$ filters used in the different WINGS-OPT observing runs, while
in Figure~\ref{fig:Filt} the transmission curves of the different
filters are shown.

With the average (dark time) observing conditions at both WFC@INT and
WFI@MPG, it turns out that the photometric depth we require for the survey
(see Section~\ref{sec:SurvReq}) can be fulfilled with exposure times
of the order of 20-25 minutes, depending on the photometric band.

In order to avoid saturation of the brightest objects, usually
three exposures per filter have been obtained, also allowing
us to easily remove cosmic rays. For A3528b (run~\#5) we have
just one exposure per filter (3m and 8m in the $B$ and $V$ band,
respectively).

We aimed for similar FWHM for each of the summed exposures. Thus, whenever
possible we tried to take these exposures with a short interval
between them. Obviously, this was not always the case for clusters
observed in service mode (runs~\#5 and \#6 with WFI@MPG). 
In particular, for nine clusters observed during the run~\#6 (A2382, 
A2399, A2717, A2734, A3667, A3716, A3809, A3880 and A4059), we got 
from ESO two medium seeing, long
exposures and a good seeing, short exposure per filter. In a forthcoming
paper of the series we will exploit this occurrence to check the
dependence of the surface photometry on the seeing.

During the first observing run we explored with a single cluster (A2107)
the possibility of taking three shifted exposures per filter in order
to fully sample the gaps between CCDs. After mosaicing, however, we
verified that, due to the worsening of the S/N ratio within the
underexposed regions, this procedure resulted in a net loss of the
area usable to perform deep surface photometry. Thus, we decided
to abandon this technique. Instead, for the whole of run~\#4, and for
many clusters observed in service mode during runs~\#5 and \#6, a
small shift in right ascension ($\sim$25pix.) was applied, allowing
us to remove bad pixels and columns.

In order to provide the WINGS-OPT survey with accurate astrometric
solutions and background galaxy counts estimation for both WFC@INT 
and WFI@MPG cameras, we have also imaged the astrometric regions 
ACR-D/E/M/N from \citet{ston} and a blank field in each
hemisphere.

Finally, some dark and dome-flat exposures and several bias frames,
twilight sky-flats and photometric standard fields have been obtained
for each observing night.  

Table~\ref{tab:ObsLog} (Online Material) reports the observing
log of the WINGS-OPT survey.

\subsection{Basic Reduction}
\label{sec:ImHand}


Most of the steps required to reduce the data coming from mosaic
wide-field cameras are similar to those usually performed on
traditional CCD frames. However, the use of such a wide area mosaic 
raises a
number of new technical issues, mainly related to the presence of
geometric distortions and photometric differences between the
different CCDs. In addition, handling the huge number of pixels from
these kind of cameras requires that even the standard reduction
procedures must be revised, to make them more efficient.  In
Appendix~\ref{sec:BasDatRed} (see Online Material) the details of the
basic reduction procedures are given.  Here we just mention that the
photometric uncertainties due to the flat fielding are expected to be
less than 1\% (0.01 mag, see Section~\ref{sec:FlatCorr}), while those
arising from bias removal and linearity correction are likely to be
negligible.  In Appendix~\ref{sec:BasDatRed} we also show that, as far
as the astrometry is concerned, the accuracy of the WINGS-OPT survey
is of the order of 0.2 arcseconds, in the worst centering situation
(big galaxies; see Section~\ref{sec:AstroCorr} and
Figure~\ref{fig:Astrom}).

\subsection{Photometric Calibration}
\label{sec:PhotCal}

Since the CCDs of any mosaic camera have usually different zero
points and color responses, the optimal standard fields for WF imaging
should map each CCD with a sufficient number of stars covering wide
ranges of both magnitude and color. For this reason, the problem of
photometric calibration in wide-field CCD mosaic cameras is not yet
solved satisfactorily. Nowadays there are two main sets of standard
fields that, even if they not provide a complete coverage of the CCD 
mosaic, can be used satisfactorily for wide field photometry, namely 
the sample of \citet{land} and that of \citet{stet}.
We preferred to use the Landolt sequences, since Stetson's standard 
fields, which go even deeper than the Landolt fields (typically
fainter than 14th magnitude, with a larger number of standard stars),
normally cover no more than 20\,arcmin on a side. 
Actually, NGC~6633
was the only Stetson standard field we used for our calibration
(run~\#4).
We used the same set of Landolt SA fields through both the INT and 
the MPG observing runs, namely SA~92/95/98/101/104/107/110/113.
During each night two or three SA fields were observed at different
zenith distances in order to map the atmospheric extinction. However,
the long average duration of each cluster pointing made it difficult
(often impossible) to observe the same standard field  more than
twice per night. In addition, the small number of
stars usually present in the standard star fields often makes 
it impossible to photometrically calibrate each CCD in a single
calibration frame.


Thus, we have performed the photometric calibration using a
self-consistent method, taking advantage of all the standard fields in
each observing run. Section~\ref{sec:PhotMethod} (Online Material)
reports both the formalism of this method and the calibration
coefficients we obtained. In particular, Figure~\ref{fig:Calib} shows,
for each observing run and for all observations of the standard stars,
the residuals (given by eq.~\ref{eq:Stand}) of our photometric
calibration in the two bands as a function of both standard magnitudes
and colors. Excluding from the calibration set the saturated and
blended stars and using a recursive $k-\sigma$ procedure to remove the
outliers, the typical $r.m.s$ of the residuals we achieved with our
calibration is of the order of $\sim$0.025~mag (see
Tables~\ref{tab:CalibRes} and ~\ref{tab:CalibRes1}).

\begin{table}[tb]
  \centering
\caption{Total $r.m.s.$ and sky transparency contribution to the $r.m.s.$ 
of the residuals of the photometric calibration in the two bands for 
each observing run of the WINGS-OPT survey.}
\label{tab:CalibRes}
  \begin{tabular}{c|cc|cc}
\hline\hline
 Run & \multicolumn{2}{|c|}{$\sigma_{\Delta B}$} & \multicolumn{2}{|c}{$\sigma_{\Delta V}$} \\
     & Total & Sky Transp. & Total & Sky Transp. \\
\hline
 \#1  & 0.020 & 0.010 &  0.017 & 0.007 \\
 \#2  & 0.023 & 0.014 &  0.018 & 0.007 \\ 
 \#4  & 0.022 & 0.009 &  0.026 & 0.014 \\
\hline
 \#3  & 0.034 & 0.020 &  0.028 & 0.016 \\              
 \#5  & 0.026 & 0.016 &  0.023 & 0.013 \\
 \#6  & 0.030 & 0.022 &  0.026 & 0.018 \\
\hline
\end{tabular}
\end{table}

To try and disentangle the different contributions to the total
$r.m.s$, we have analysed different nights of the same run. In
Table~\ref{tab:CalibRes}, the right column relative to each filter
reports the contribution to the scatter arising from sky transparency
fluctuations through the run. In particular, the night-, run- and
long-term contributions to these fluctuations, estimated normalizing
the residuals relative to each individual star to their night-, run-
and long-term averaged values, respectively, are found to be roughly
equivalent among each other. However, from Table~\ref{tab:CalibRes} it
is clear that the different contributions due to sky transparency
variations, altogether, do not represent the dominant share of the
scatter in the photometric calibration. This is likely due to
systematic effects arising from both possible zero point gradients
across the fields and differences among the photometric systems.

\begin{figure}
   \vspace{-0.5cm}
   \hspace{-0.5cm}	
    \includegraphics[width=10cm]{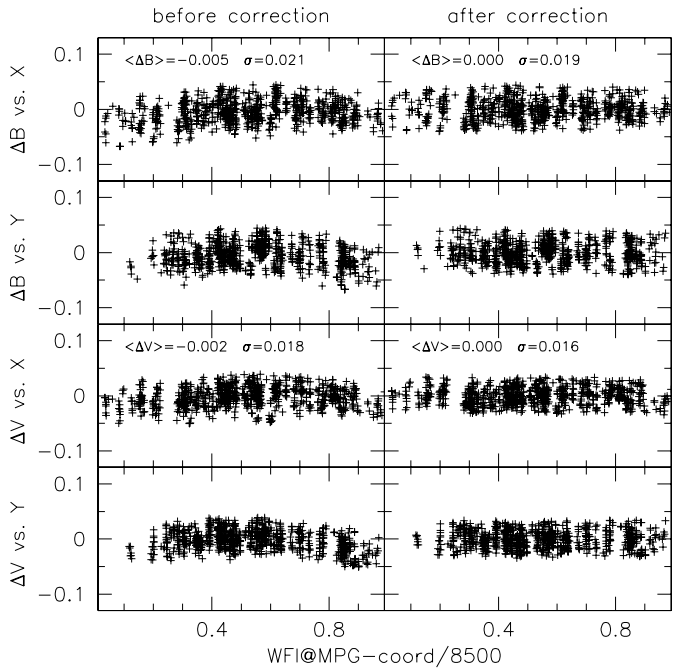}
   \vspace{-0.5cm}
  \caption{Residuals of our calibration versus the
pixel coordinates, for the whole set of standard stars observed with WFI@MPG,
before (left panels) and after (right panels) correction}
  \label{fig:ZPmap}
\end{figure}

Concerning the former effect, in Appendix~\ref{sec:BasDatRed}
(Section~\ref{sec:FlatCorr}) we report on the non-uniform illumination
of the imaging taken with WFI@MPG, which can induce systematic
magnitude differences up to $\sim$0.1~mag across the field. Even
though our chip by chip photometric calibration procedure (see
Table~\ref{tab:Calib1} in Appendix~\ref{sec:PhotCalProc}) should in
principle alleviate this problem, we have directly verified the
non-uniformity of our photometric zero points by plotting in
Figure~\ref{fig:ZPmap} (left panels) the residuals of our calibration
versus the pixel coordinates for the whole set of standard stars
observed with WFI@MPG. Since in both filters a significant dependence
on the position is found to persist for the residuals, we have
interpolated them through the field using a 2nd-order, 2D polynomial.
The right-hand panels of Figure~\ref{fig:ZPmap} show that the residuals, 
after correction, no longer depend on the position.

No significant spatial gradients of the residuals of the photometric
calibration were found in the case of the WFC@INT camera.
Table~\ref{tab:CalibRes1} summarizes the different 
contributions to the scatter, averaged over the whole data-set of
standard stars observations available for each camera.

\begin{table}[tb]
  \centering
\caption{
Different contributions to the $r.m.s.$ of the residuals of the 
photometric calibration in the two bands and for each WF camera
}
\label{tab:CalibRes1}
  \begin{tabular}{l|cc|cc}
\hline\hline
 Contribution & \multicolumn{2}{|c|}{WFC@INT} & \multicolumn{2}{|c}{WFI@MPG} \\
              & $\sigma_{\Delta B}$&$\sigma_{\Delta V}$&$\sigma_{\Delta B}$&$\sigma_{\Delta V}$\\
\hline
 Sky Transp. &  0.013   &  0.011   &  0.019   &  0.015   \\
 Phot.Syst. &  0.018   &  0.020   &  0.018   &  0.016   \\
 ZP Gradient    &    -     &    -     &  0.010   &  0.010   \\
\hline
 Total       &  0.022   &  0.023   &  0.028   &  0.024   \\
\hline
\end{tabular}
\end{table}

\subsection{Mosaics}
\label{sec:Mosaic}

After having gone through the usual reduction steps (de-biasing,
linearity correction, flat-fielding, astrometry), the multi-extension
exposures of each given cluster in each filter have been registered,
co-added and mosaiced using the \verb; wfpred; package 
(see Appendix~\ref{sec:BasDatRed}). 

Figures~\ref{fig:MosaicINT} and \ref{fig:MosaicMPG} show examples
of the mosaic imaging obtained with the WFC@INT and WFI@MPG
cameras, respectively.
We produced co-added and mosaiced frames even when the different
exposures of a given cluster came from different observing nights,
with different observing conditions. However, in these cases, the
mosaics of just the exposures with comparable conditions were
produced. For instance, when two medium seeing, long exposures and a
good seeing, short exposure were available in each filter (five clusters
observed during run~\#6; see Section~\ref{sec:Obs}), besides the
co-added mosaic of the three exposures, we produced that of the two
medium seeing exposures and the mosaic of the good seeing exposure. In
fact, each one of them could be suitable for a particular task
(integrated photometry, surface photometry, morphology).

\begin{figure*}
\begin{center}
\includegraphics[width=14cm]{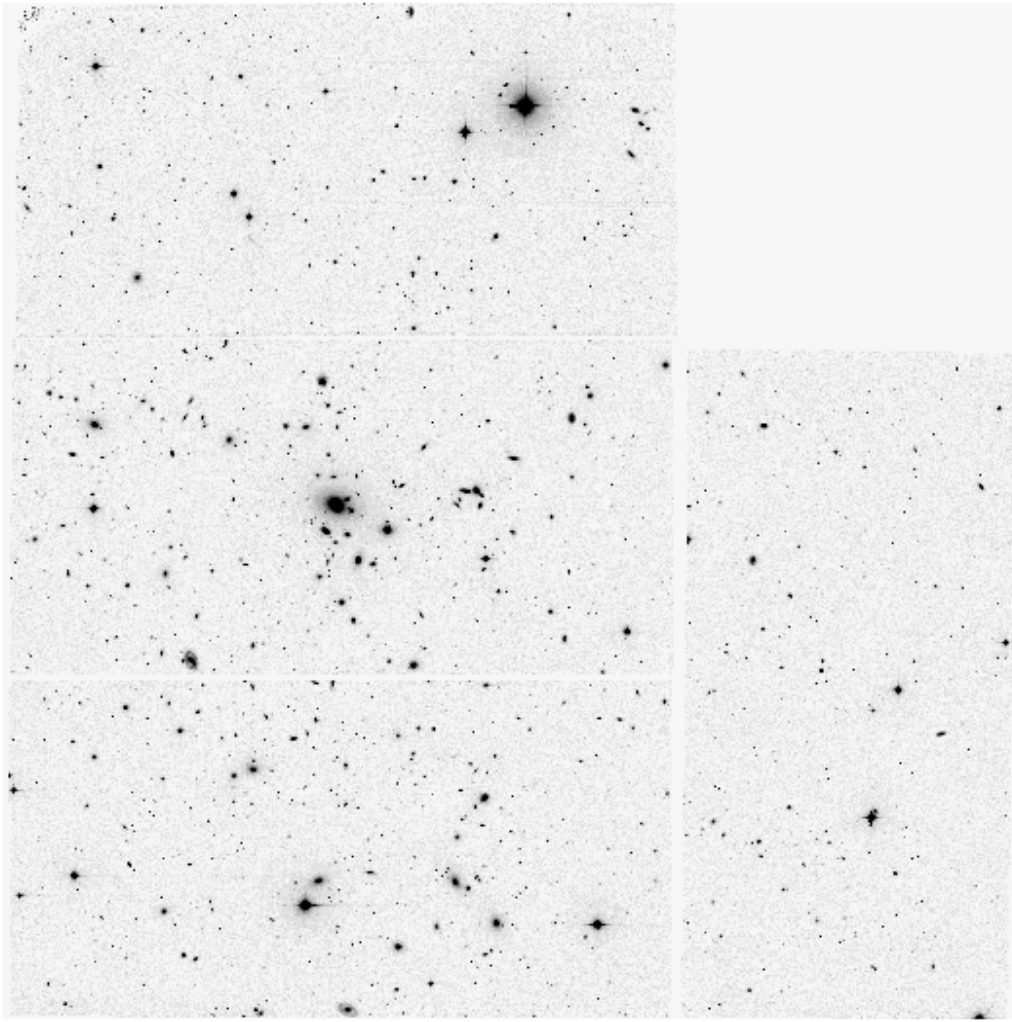}
\caption{Mosaic of the WFC@INT image of the cluster A151. North is up, East is left.
The field of view is $34^\prime\times 34^\prime$.}
\label{fig:MosaicINT}
\end{center}
\end{figure*}

\begin{figure*}
\begin{center}
\includegraphics[width=19cm]{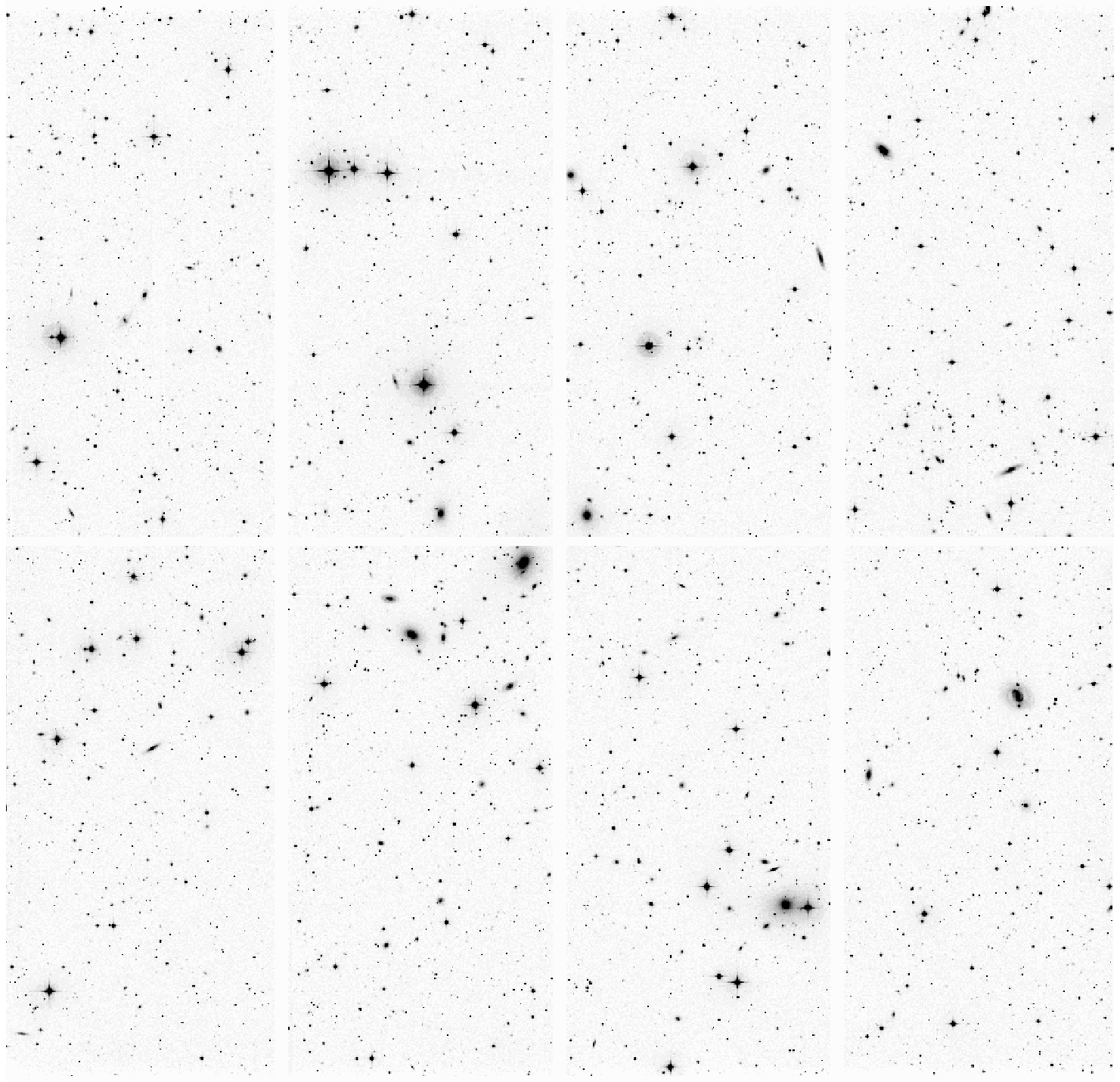}
\caption{Mosaic of the WFI@MPG image of the cluster A3556. North is up, East is left.
The field of view is $34^\prime\times 33^\prime$.}
\label{fig:MosaicMPG}
\end{center}
\end{figure*}

Before extracting the photometric quantities to be included in our
catalogs (Section~\ref{sec:Catalogues}), we have put the co-added mosaic
frames through a normalization procedure accounting for the different
photometric coefficients of the mosaic's CCDs. This procedure is
described in Section~\ref{sec:MosaicProc} (Online Material).

\subsection{Cosmetics}
\label{sec:Cosmetics}

Since for each cluster three exposures, with a short
interval between them, were usually obtained for each filter (see
Section~\ref{sec:Obser}), the co-adding procedure was in general
sufficient to remove cosmic rays. When less than three close exposures
were available, we resorted to the IRAF tool COSMICRAYS to do the job.

For nearly half of the cluster sample (run~\#4 and part of runs~\#5
and \#6) the three available exposures were dithered by $\sim$25
pixels, allowing us to remove the bad pixels and columns. For the
remaining clusters, pixel mask images were automatically produced and
used by the IMEDIT-IRAF tool to interpolate the bad regions. We were
forced to adopt this technique because of the noticeable worsening of
the photometric accuracy we found in the experiments carried out with
SExtractor \citep{bert} when weight-images are used to account for bad pixels and
columns.

\subsection{Background removal}
\label{sec:BackRem}

Estimating the local background is a crucial step in achieving
good quality photometry. In our case, the main problems related to
the background removal reside in the presence of objects with extended
halos (big early-type galaxies) or wings (very bright stars), as well
as in the discontinuity of the background associated with the gaps
between different CCDs. Both are likely to produce 
artificial distortions in the background map, thus systematically 
biasing the local backgrond estimates.

We exploited the capabilities of SExtractor, as well as the
ELLIPSE-IRAF tool to devise a semi-automatic, iterative procedure for
optimal sky subtraction over CCD mosaics, even in case of crowded
galaxy cluster fieds, possibly including big halo galaxies and/or very
bright stars. This procedure generates two images. The first is the
original mosaic, after model subtraction of the big halo galaxies and
very bright stars. The second image contains only the previously
removed big/bright galaxies, where the masked pixels (neighbours or
gaps) are replaced by the models. These two images are suitable for
SExtractor processing, since each one of them contains homogeneously
sized objects, without critical blendings.

\subsection{Catalogues}
\label{sec:Catalogues}

\begin{figure}
    \includegraphics[width=9cm]{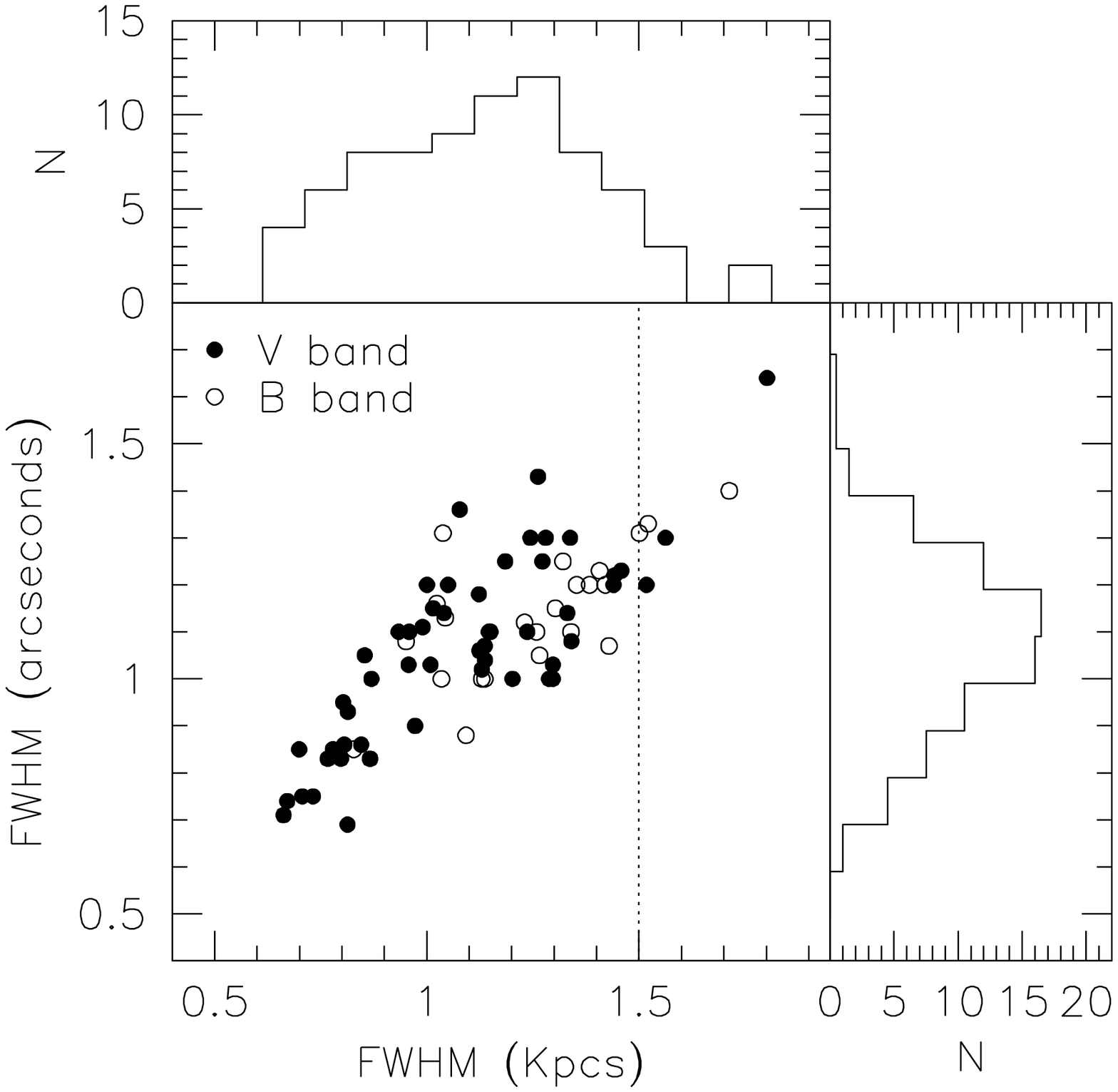}
  \caption{The FWHM in arcseconds versus the physical size this
projects to at the cluster redshift, with the marginal distributions 
for each, for the WINGS-OPT imaging. Filled and open circles indicate 
that the best FWHM is achieved in the $V$ and $B$ band, respectively.}
  \label{fig:fwresol}
\end{figure}

\begin{figure}
    \includegraphics[width=9cm]{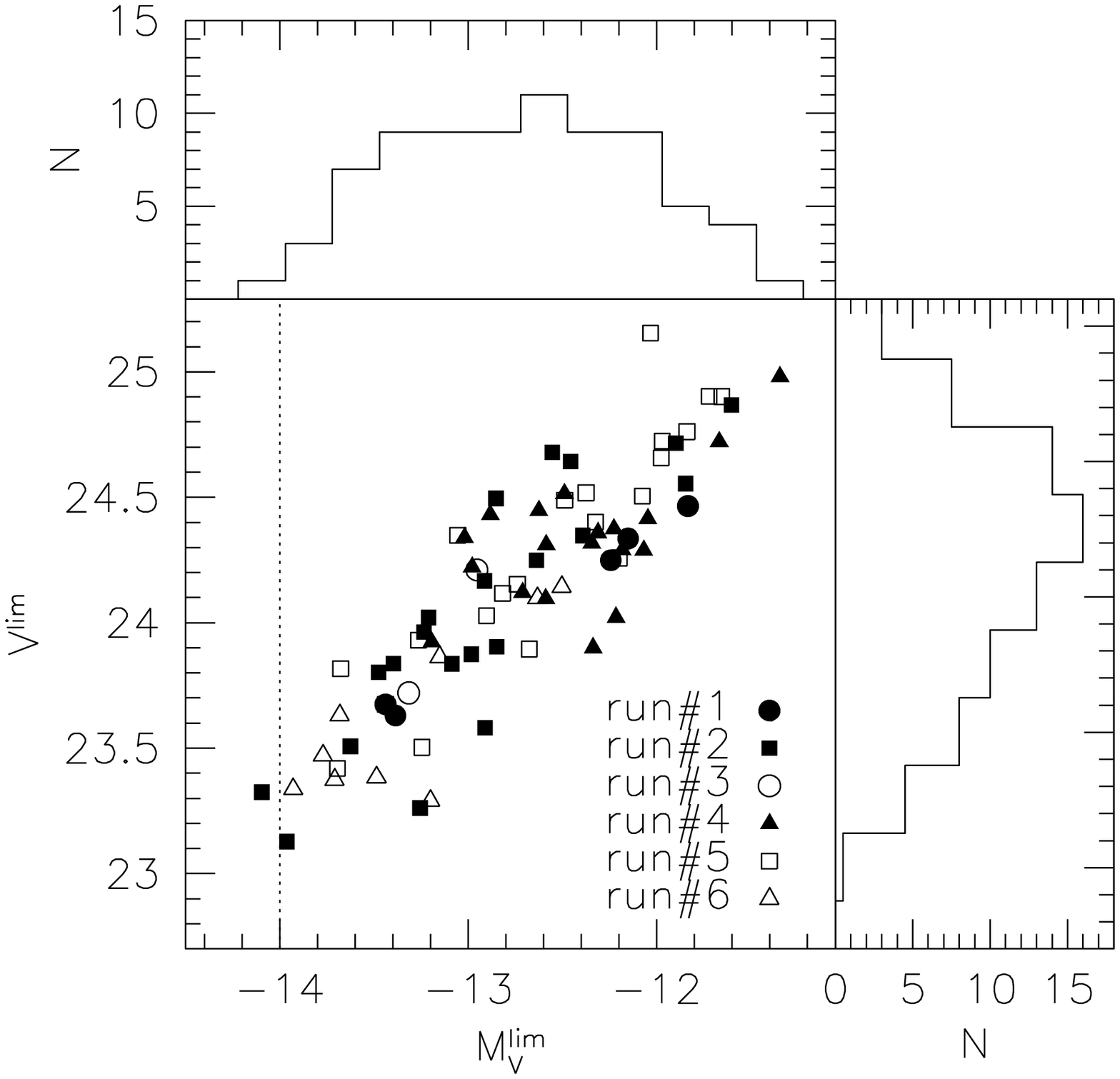}
  \caption{Apparent versus absolute $V$-band magnitudes at the detection limit, 
with the marginal distributions for each, for our WINGS-OPT observations. 
Different observing runs are plotted with different symbols: 
full and open symbols referring to the WFC@INT and WFI@MPG observations, respectively.}
  \label{fig:DetecLim}
\end{figure}

The final photometric catalogs of the WINGS-OPT survey are obtained,
for each cluster, by running SExtractor over the two previously
described images in both wavebands and by merging the four resulting
catalogs into a single master-catalog containing all the sources detected in
both filters over the field.
The magnitudes in the final catalogs are color corrected following the
procedure outlined in Section~\ref{sec:MosaicProc} (equations~\ref{eq:ColCorr} 
and \ref{eq:ColCorr1}). 

At this stage, we tried to detect as many sources as possible by adopting
very liberal detection parameters within SExtractor. In particular, we used 
a minimum detection area of 5 pixels and detection thresholds of 1.5 
and 1.1 times the $\sigma_{bkg}$ for the WFC@INT and WFI@MPG imaging, 
respectively, roughly corresponding to S/N$\approx$4.5 per square 
arcseconds in both cases.

In a forthcoming paper of the WINGS series, we will present the
WINGS-OPT catalogs, describing in detail the procedure we used to
produce them. Here we just
mention that, on the basis of the automatic
star/galaxy classifier (S/G) given by SExtractor, the master catalog
for each cluster was split into three preliminary catalogs: (i) a
galaxy catalog (GCAT, S/G$\leq$0.2); (ii) a star catalog (SCAT,
S/G$\geq$0.8); (iii) a catalog of objects with uncertain
classification (UCAT, 0.2$<$S/G$<$0.8). Finally, with the use
of the multi-aperture photometry plotting tools and our
visual inpsections of the final images, the catalogs
are carefully cleaned, with spurious detections (residual spikes and bad pixels, border
effects, etc..) removed and mis-classified objects moved 
from one catalog to another (GCAT into SCAT and viceversa).

\section{The WINGS-OPT data quality}
\label{sec:Quality}

In Section~\ref{sec:SurvReq} we set the minimal requirements that the WINGS-OPT
imaging survey should obey as far as both the spatial
resolution (FWHM$\leq$1.5~Kpc) and the limiting absolute
magnitude ($M_V^{lim}\geq -14$) are concerned. Using the photometric
catalogues we have checked to what extent these requirements have
been fulfilled by the actual WINGS-OPT data.

\subsection{Spatial resolution}
\label{sec:CatSpaRes}

In Figure~\ref{fig:fwresol} the FWHM in arcseconds is plotted
against the actual physical resolution this projected to at
the redshift of the cluster (expressed in kpc), for all our 
WINGS-OPT observations. Apart from a few very bad cases, the bulk of our
cluster sample, in both arcseconds and kiloparsecs, is located around
FWHM$\sim$1.1 (see histograms in the figure). In spite of the repeated
observations taken in different runs, the spatial resolution of two
clusters (A1668, A2626) largely exceeds the requirement described in
Section~\ref{sec:SurvReq}. These clusters will be flagged out in the
statistical analyses of surface photometry and morphology results.

\subsection{Photometric depth}
\label{sec:CatDeep}

\begin{figure}
    \includegraphics[width=9cm]{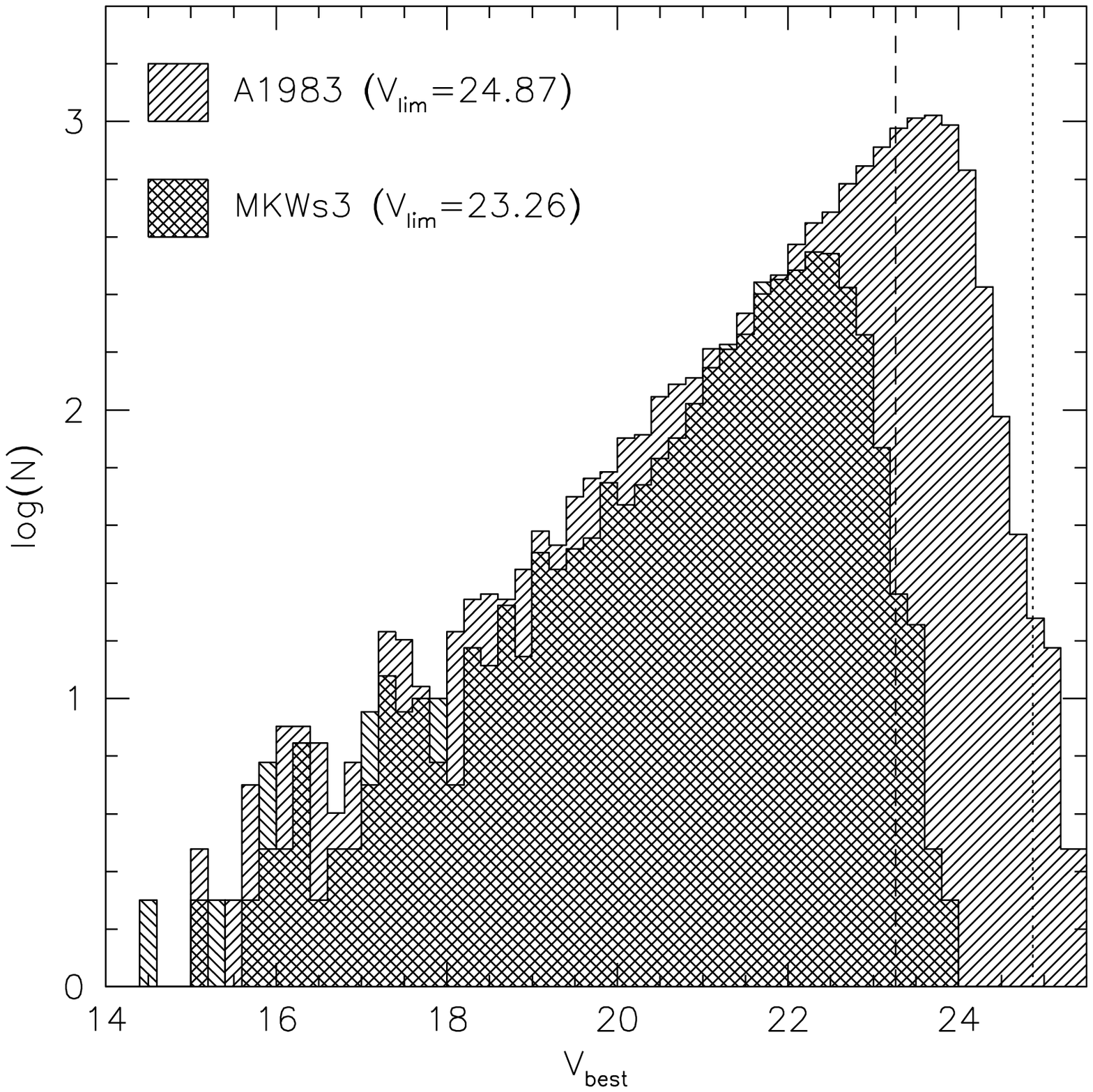}
  \caption{V-band magnitude histograms for A1983 and MKW3s, compared with the
corresponding detection limit magnitudes computed using 
equation~\ref{eq:DetecLim}(vertical lines).}
  \label{fig:HisMag}
\end{figure}

As far as the photometric depth of the survey is concerned, during
observing run~\#2, fourteen clusters were observed in good seeing but
in uncertain photometric conditions, and so were imaged again in good
photometric conditions. The photometric, short exposures were used to
calibrate the long exposures with uncertain photometry. For eight of
these clusters, the photometric adjustments turned out to be negligible
in both filters, while for two clusters (A970 and A1069), a correction
of $\sim$0.18~mag was needed in the B band only. The comparison with 
photometric exposures did however show that large corrections
(from 0.6~mag to 1.2~mag) in both bands were needed for the three
clusters A2149, A2271 and MKW3s, whose photometric depth turned out 
to be irreparably worsened. 

In spite of this, it turns out from Figure~\ref{fig:DetecLim} that the
requested minimal absolute depth was achieved for practically all
clusters in the WINGS-OPT survey. The only case where $M_V^{lim}$ 
exceeded the requested limit, A2149, it did so by just a few hundredths of
a magnitude. The detection
limits reported in Figure~\ref{fig:DetecLim} are computed using the formula:
\begin{equation}
m_{lim}=Z-2.5\times\log(\pi\times\nu\times\sigma_{bkg}\times{\rm FWHM}^2_{pix}),
\label{eq:DetecLim}
\end{equation}
where $Z$ is the photometric zero point, $\sigma_{bkg}$ is the
$r.m.s.$ of the background and $\nu$ is the detection threshold 
in units of $\sigma_{bkg}$. in Figure~\ref{fig:DetecLim} we set
$\nu$=1.5 and $\nu$=1.1 for WFC@INT and WFI@MPG observations, respectively
(see Section~\ref{sec:Catalogues}). 
It is worth noting that these `theoretical' detection
limits turn out to be in fair agreement with the
actual limits derived from the master catalogs for our 
clusters. This is well illustrated in Figure~\ref{fig:HisMag}, where
the $V$-band magnitude histograms for two clusters (A1983 and MKW3s), 
representing extreme cases of different photometric depth, are
compared with the detection limit magnitudes computed using
equation~\ref{eq:DetecLim}. From Figure~\ref{fig:HisMag} it turns out
that, even if there are detections well beyond the `theoretical' limit, the
actual completeness limit achieved is approximately half magnitude brighter
than this limit.

We also note that, the average detection limits of the WFC@INT
and WFI@MPG observations turned out to be similar
($<V^{lim}>\sim$24.1) -- as expected from the exposure time
calculators --  while the corresponding average
completeness magnitude of the survey is $V^{comp}\sim$23.5.

\subsection{Internal photometric consistency}
\label{sec:CatIntPhotCons}


To check the internal consistency of the photometry given in the
WINGS-OPT master catalogs, determining at the same time how the
photometric random errors depend on the flux, we compare in 
Figure~\ref{fig:CompPhot1} the magnitudes of stars in those clusters which
have been observed on different nights during the same run, or during
different runs with the same camera, or even with different
cameras. Left, middle and right panels in Figure~\ref{fig:CompPhot1}
show the magnitude differences as a function of the (average)
magnitude itself for INT-INT, MPG-MPG and INT-MPG comparisons,
respectively. Bottom panels of the same figure show the behaviour
of the observed $r.m.s.$ due to random errors as a function of
magnitude (dots), compared with the expected
theoretical functions (dashed lines), computed using the proper,
specific observational parameters.

The systematic magnitude shifts in Figure~\ref{fig:CompPhot1} are
generally consistent with the expected zero point fluctuations
among different observations (see Table~\ref{tab:CalibRes1}).
Also the random errors turn out to be in fair agreement with the 
expectations, apart from the INT-MPG comparison, where an additional
source of scatter is present.

\subsection{External photometric consistency}
\label{sec:CatExtPhotCons}

\begin{figure*}
\vskip -4.5truecm	
\hspace{-2cm}
\includegraphics[width=21.5cm]{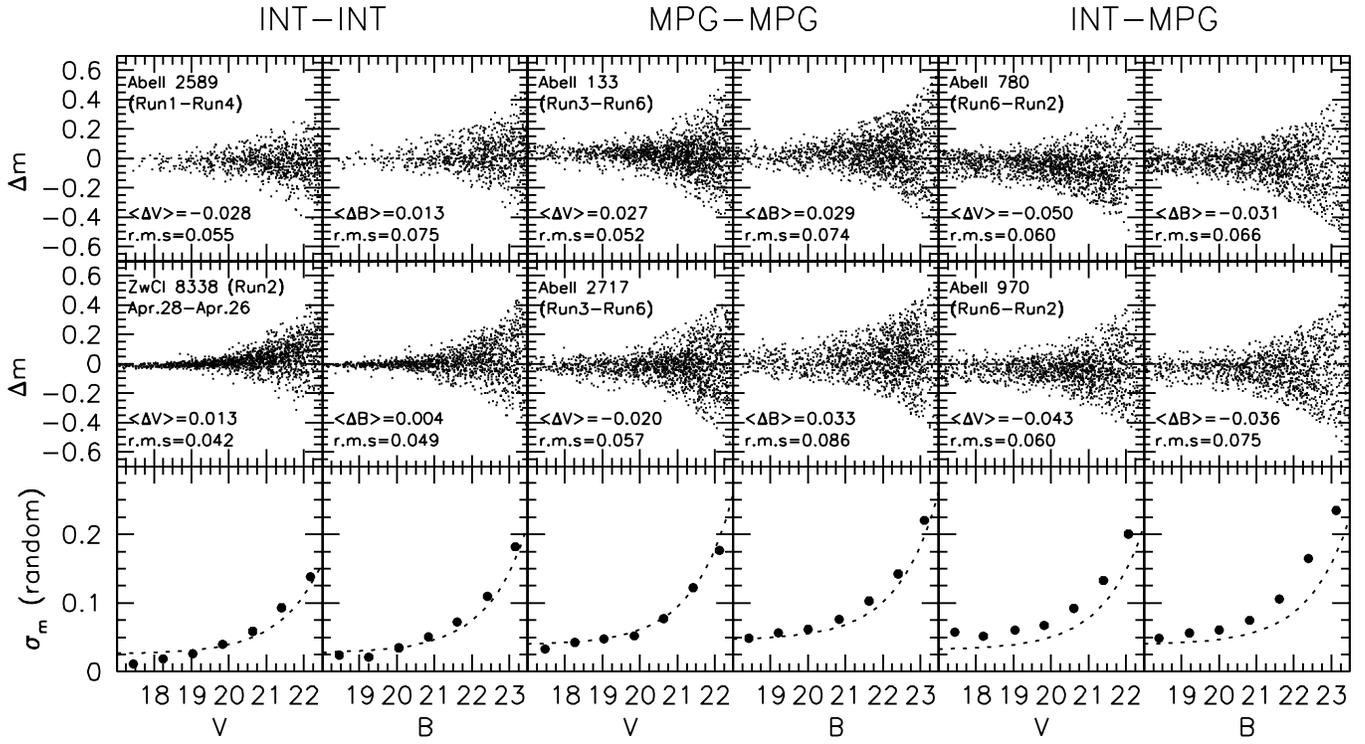}
\vskip -6truecm
\caption{Magnitude differences as a function of
the (average) magnitude itself for INT-INT (leftmost two panels), 
MPG-MPG (central panels) and INT-MPG (rightmost two panels)
comparisons in some clusters which have been observed on different
nights during the same run, or during different runs with the same camera, or
even with different cameras. Bottom panels show
the behaviour of the observed $r.m.s.$ due to random errors as a
funtion of the magnitude (dots), compared with the
expected theoretical functions (dashed lines), computed using the
proper, specific observational parameters.}
\label{fig:CompPhot1}
\end{figure*}

In order to perform an external consistency check of our photometric 
system, we have compared the magnitudes of stars in our master catalogs
of Abell~119 (North) and Abell~2399 (South) with those provided for 
the same fields by the SDSS Sky Server.
In the upper panels of Figure~\ref{fig:CompSDSS} the star magnitude differences
$V_{WINGS}-r'_{SDSS}$ are reported as a function of the colors 
$(B-V)_{WINGS}$ and these color-color plots are compared with the conversion
equation (23) in \citet{fuku}. In this figure just the stars brighter 
than $V=20$ are reported.
The lower panels of Figure~\ref{fig:CompSDSS} show, as a function of
the magnitude, the differences between our $V$ magnitudes and the
corresponding SDSS magnitudes, derived using the above mentioned equation.

The agreement between the two photometric systems turns out to be quite good
and the random scatter as a function of magnitude looks quite similar 
to that found in the case of the internal consistency check 
(see Figure~\ref{fig:CompPhot1}).

\subsection{Overall photometric quality}
\label{sec:CatPhotAcc}

\begin{figure}
\hspace{-0.5cm}
\includegraphics[width=10cm]{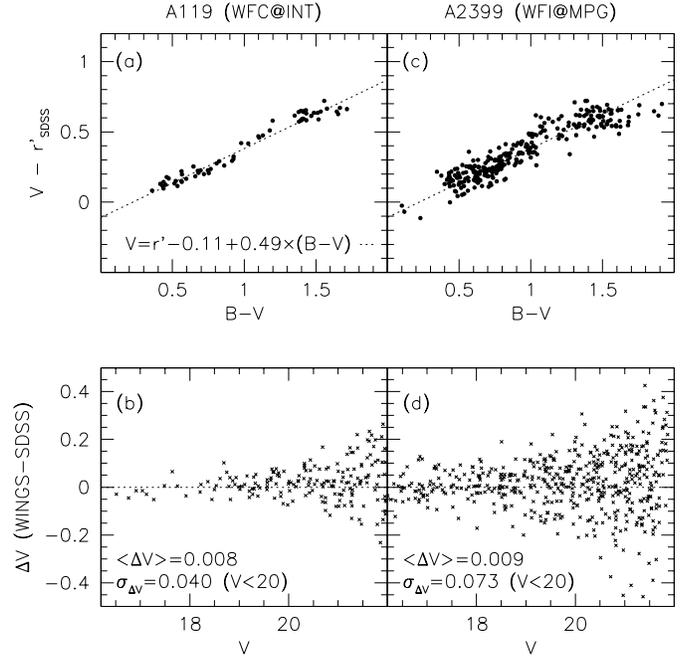}
\caption{Panels~(a) and (c): $V_{WINGS}-r'_{SDSS}$ as a function of $(B-V)$ for
the stars brighter than $V=20$ in the fields of Abell~119 (WFC@INT) and Abell~2399
(WFI@MPG). The dotted lines represent the color conversion proposed by \citet[ eq.(23)]{fuku}. 
Panels~(b) and (d): comparison between WINGS and SDSS
magnitudes, after conversion of the r'(SDSS) magnitudes into the $V$
band using the equation by \citet{fuku}.}
\label{fig:CompSDSS}
\end{figure}

\begin{figure}
    \includegraphics[width=9cm]{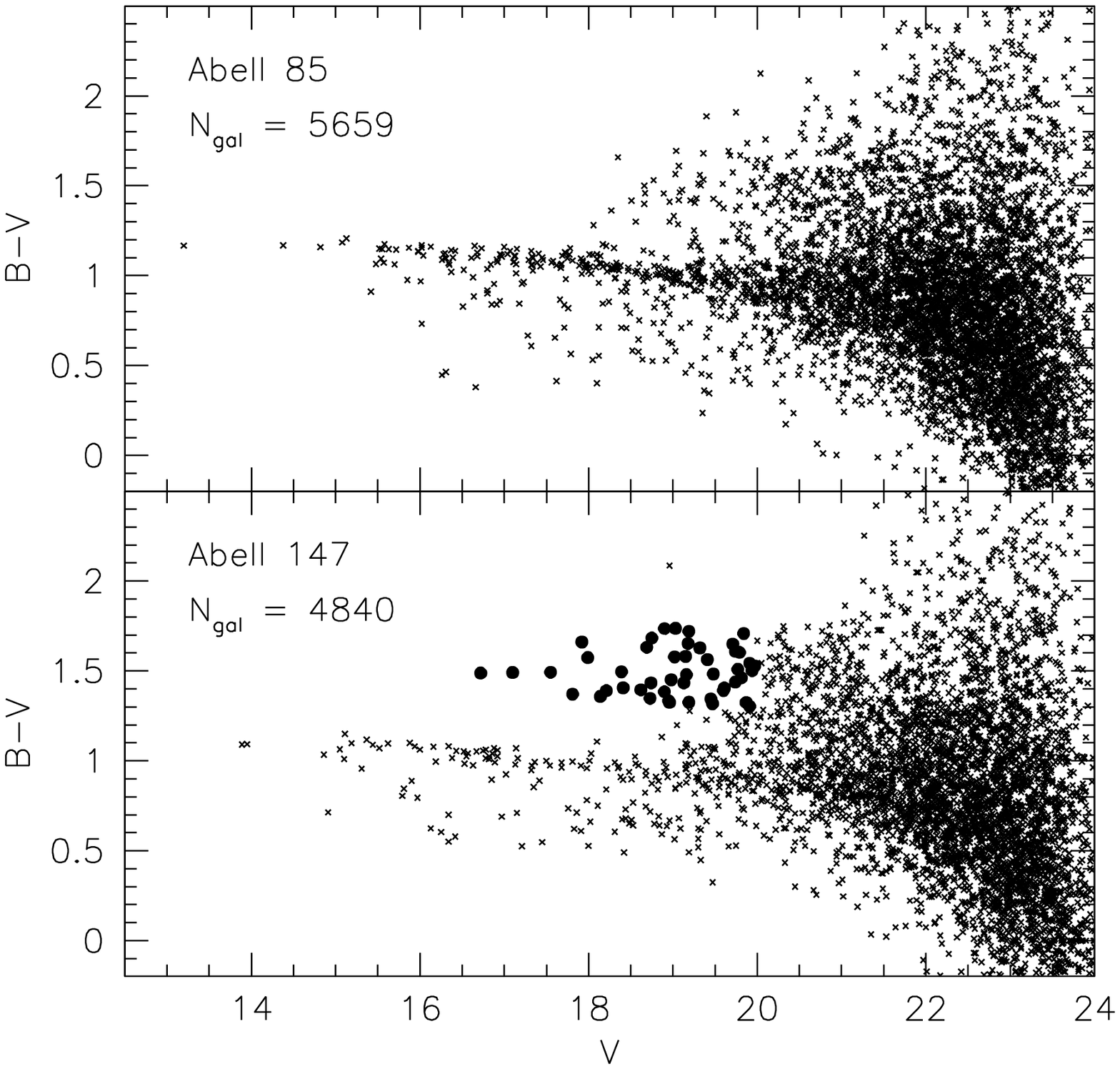}
  \caption{Color-Magnitude diagrams from the WINGS-OPT master catalogs
for the clusters Abell~85 and Abell~147. Note the second redder sequence
in Abell~147 (full dots), which is likely to indicate the presence of
a second galaxy cluster/group in the background}
  \label{fig:CM}
\end{figure}

From Table~\ref{tab:CalibRes1} and from Figures~\ref{fig:CompPhot1} 
and \ref{fig:CompSDSS} we conclude that the total (systematic plus 
random) photometric $r.m.s.$ errors of our survey,
derived by both internal and external comparisons vary from
$\sim$0.02~mag, for bright objects, up to $\sim$0.2~mag, for objects
close to the detection limit. However, it is worth noting that,
since the above analysis is based on magnitudes derived
by SExtractor, it refers mainly to point
sources. Actually, the systematic errors involved in the estimation of
total galaxy magnitudes are known to depend on the galaxy light
profile, as well as on the average surface brightness of galaxies
\citep{fran}. In a forthcoming paper of the series we
will perform this analysis in the specific case of the WINGS-OPT
survey. However, to illustrate the photometric quality of our galaxy
dataset, even in the preliminary form provided by the WINGS-OPT
master catalogs, we show in Figure~\ref{fig:CM} two examples of the
color-magnitude relations derived for our WINGS clusters.

\section{Summary and future plans}
\label{sec:FutPlans}

The WINGS-OPT observations we have presented here are part of an
ambitious project aimed at providing the astronomical community with a
huge database of galaxy properties in nearby clusters, to be used as a
local benchmark for evolutionary studies.

We have described in detail our optical imaging, as well as the
reduction procedures we used to manage the different issues associated
with the wide-field mosaics. All the steps of the reduction sequence have
been carefully checked for correspondence between expected and actual
results and special care has been paid to control the quality of
astrometry and photometry. As far as the first issue is concerned, the
typical $r.m.s.$ of the astrometric errors is found to be of the order
of 0.2 arcseconds in both the northern and the southern
observations. The photometric quality has been controlled using both
internal and external consistency checks. In both cases the average
differences among different observations turn out to be of the order
of a few hundredths of magnitude, while the random
photometric errors ($r.m.s.$) increase with increasing magnitude, from
$\sim$0.02~mag for bright objects up to $\sim$0.2~mag for objects
close to the detection limits. These limits are $\sim$24~mag
and $\sim$25~mag in the $V$ and $B$ bands, respectively, allowing us to sample
the luminosity function of galaxies down to $M_V\sim$-14 for almost
all clusters and down to $M_V\sim$-13 for roughly half the sample.

We have also checked {\it a posteriori} whether the global quality of
the WINGS-OPT imaging is actually consistent with the minimum standards we
set {\it a priori}. We found that only for a few
clusters the actual image quality (in terms of seeing and photometric depth)
turns out to be marginally worse than the formal requirements.

The catalogs used to perform the above analyses have been produced for
each cluster by running SExtractor on the mosaiced frames in both
filters. They contain position, shape and photometry parameters of
several thousands of stars and galaxies in the cluster field. In this
paper we have just outlined the complex procedure used to produce the
catalogs. In a forthcoming paper we will go into more detail about
catalogs, making them available for the whole astronomical
community. Subsequent papers of the series will concern surface photometry and
morphological classification of a subsample of large galaxies (more
than 200~pix above 1.5$\sigma_{bkg}$), the global
cluster properties (total luminosity and luminosity profile,
characteristic radius, flattening) and the analysis of
subclustering. Later, we will concentrate on the
statistical properties of galaxies (luminosity function,
color-magnitude, $\muere$ and morphology-density relations) as a
function of both the cluster properties and the environment (position
inside the cluster and local density). In parallel, we also plan
to produce the spectroscopic database, including redshifts and line 
indices of brightest galaxies, about 100 to 300 per cluster.

\begin{acknowledgements}

We wish to thank E.V.~Held for the helpfull discussions on
treatment of wide-field data.

This work has been partially funded by the Italian Ministry of
Education and Reasearch (MIUR) through the project
N$^\circ$.~2001021149 (2002).

\end{acknowledgements}

\bibliographystyle{aa}
\bibliography{WINGS-I}

\Online

\begin{longtable}{lcrccccc}
\caption{\label{tab:Sample} The WINGS cluster sample.}\\
\hline\hline
Cluster & $\alpha$ & \multicolumn{1}{c}{$\delta$} & Redshift & Abell & B-M & $L_X/10^{44}$ & E(B-V) \\
& \multicolumn{2}{c}{(J2000)} & & Rich. & Type & ergs~s$^{-1}$ & mag \\
\hline
\endfirsthead
\caption{continued.}\\
\hline\hline
Cluster & $\alpha$ & \multicolumn{1}{c}{$\delta$} & Redshift & Abell & B-M & $L_X/10^{44}$ & E(B-V) \\
& \multicolumn{2}{c}{(J2000)} & & Rich. & Type & ergs~s$^{-1}$ & mag \\
\hline
\endhead
\hline
A0085    & 00~41~50 & -09~18 & 0.0521 & 1 &   I    &   4.27 &  0.038 \\
A0119    & 00~56~21 & -01~15 & 0.0444 & 1 & II-III &   1.65 &  0.038 \\
A0133    & 01~02~42 & -21~52 & 0.0603 & 0 &    -   &   1.82 &  0.018 \\
A0147    & 01~08~12 &  02~11 & 0.0447 & 0 &  III   &   0.28 &  0.025 \\
A0151    & 01~08~51 & -15~24 & 0.0536 & 1 &   II   &   0.52 &  0.026 \\
A0160    & 01~13~04 &  15~30 & 0.0442 & 0 &  III   &   0.19 &  0.086 \\
A0168    & 01~15~09 &  00~17 & 0.0448 & 2 & II-III &   0.56 &  0.035 \\
A0193    & 01~25~07 &  08~41 & 0.0485 & 1 &   II   &   0.78 &  0.051 \\
A0311    & 02~09~28 &  19~46 & 0.0657 & 0 &    -   &   0.41 &  0.174 \\
A0376    & 02~46~04 &  36~54 & 0.0488 & 0 &  I-II  &   0.71 &  0.073 \\
A0500    & 04~38~52 & -22~06 & 0.0670 & 1 &  III   &   0.72 &  0.050 \\
A0548b   & 05~45~28 & -25~55 & 0.0441 & 1 &  III   &   0.15 &  0.029 \\
A0602    & 07~53~26 &  29~21 & 0.0621 & 0 &  III   &   0.57 &  0.057 \\
A0671    & 08~28~32 &  30~25 & 0.0503 & 0 & II-III &   0.45 &  0.047 \\
A0754    & 09~08~32 & -09~37 & 0.0542 & 2 &  I-II  &   4.08 &  0.064 \\
A0780    & 09~18~06 & -12~05 & 0.0565 & 0 &    -   &   3.38 &  0.045 \\
A0957    & 10~13~38 & -00~55 & 0.0448 & 1 &  I-II  &   0.40 &  0.042 \\
A0970    & 10~17~34 & -10~40 & 0.0595 & 1 &  III   &   0.77 &  0.054 \\
A1069    & 10~39~43 & -08~41 & 0.0622 & 0 &  III   &   0.48 &  0.041 \\
A1291    & 11~32~21 &  55~58 & 0.0527 & 1 &  III   &   0.22 &  0.019 \\
A1631a   & 12~52~52 & -15~24 & 0.0466 & 0 &   I    &   0.37 &  0.054 \\
A1644    & 12~57~11 & -17~24 & 0.0475 & 1 &   II   &   1.80 &  0.072 \\
A1668    & 13~03~46 &  19~16 & 0.0634 & 1 &   II   &   0.81 &  0.032 \\
A1736    & 13~27~11 & -27~12 & 0.0461 & - &  III   &   1.21 &  0.058 \\
A1795    & 13~48~52 &  26~35 & 0.0622 & 2 &   I    &   5.67 &  0.013 \\
A1831    & 13~59~15 &  27~58 & 0.0612 & 1 &  III   &   0.97 &  0.019 \\
A1983    & 14~52~59 &  16~42 & 0.0444 & 1 &  III   &   0.24 &  0.027 \\
A1991    & 14~54~31 &  18~38 & 0.0586 & 1 &   I    &   0.69 &  0.033 \\
A2107    & 15~39~39 &  21~46 & 0.0411 & 1 &   I    &   0.56 &  0.057 \\
A2124    & 15~44~59 &  36~06 & 0.0654 & 1 &   I    &   0.69 &  0.025 \\
A2149    & 16~01~35 &  53~55 & 0.0675 & 0 &    -   &   0.42 &  0.010 \\
A2169    & 16~14~09 &  49~09 & 0.0579 & 0 &  III   &   0.23 &  0.015 \\
A2256    & 17~03~35 &  78~38 & 0.0581 & 2 & II-III &   3.60 &  0.053 \\
A2271    & 17~18~17 &  78~01 & 0.0584 & 0 &   I    &   0.32 &  0.042 \\
A2382    & 21~51~55 & -15~42 & 0.0644 & 1 & II-III &   0.46 &  0.057 \\
A2399    & 21~57~13 & -07~50 & 0.0582 & 1 &  III   &   0.51 &  0.039 \\
A2415    & 22~05~40 & -05~36 & 0.0590 & 0 &  III   &   0.86 &  0.066 \\
A2457    & 22~35~41 &  01~29 & 0.0591 & 1 &  I-II  &   0.73 &  0.084 \\
A2572a   & 23~17~13 &  18~42 & 0.0422 & 0 &  III   &   0.52 &  0.051 \\
A2589    & 23~23~57 &  16~46 & 0.0416 & 0 &   I    &   0.95 &  0.030 \\
A2593    & 23~24~20 &  14~38 & 0.0428 & 0 &   II   &   0.59 &  0.044 \\
A2622    & 23~35~01 &  27~22 & 0.0613 & 0 & II-III &   0.55 &  0.057 \\
A2626    & 23~36~30 &  21~08 & 0.0565 & 0 &  I-II  &   0.99 &  0.063 \\
A2657    & 23~44~57 &  09~11 & 0.0400 & 1 &  III   &   0.82 &  0.126 \\
A2665    & 23~50~50 &  06~09 & 0.0562 & 0 &    -   &   0.97 &  0.080 \\
A2717    & 00~03~13 & -35~56 & 0.0498 & 0 &  I-II  &   0.52 &  0.011 \\
A2734    & 00~11~22 & -28~51 & 0.0624 & 0 &  III   &   1.30 &  0.017 \\
A3128    & 03~30~15 & -52~32 & 0.0590 & - &  I-II  &   1.08 &  0.016 \\
A3158    & 03~43~09 & -53~39 & 0.0590 & - &  I-II  &   2.71 &  0.015 \\
A3164    & 03~45~49 & -57~02 & 0.0611 & 0 &  I-II  &   0.75 &  0.027 \\
A3266    & 04~31~13 & -61~27 & 0.0545 & - &  I-II  &   3.14 &  0.020 \\
A3376    & 06~00~41 & -40~02 & 0.0464 & 0 &   I    &   1.27 &  0.052 \\
A3395    & 06~27~36 & -54~26 & 0.0497 & 0 &   II   &   1.43 &  0.113 \\
A3490    & 11~45~20 & -34~26 & 0.0697 & 2 &   I    &   0.88 &  0.087 \\
A3497    & 12~00~04 & -31~23 & 0.0609 & 0 &  I-II  &   0.74 &  0.072 \\
A3528a   & 12~54~35 & -29~23 & 0.0535 & 0 &   II   &   0.68 &  0.078 \\
A3528b   & 12~54~00 & -28~51 & 0.0535 & 0 &    -   &   1.01 &  0.078 \\
A3530    & 12~55~36 & -30~20 & 0.0544 & 0 &  I-II  &   0.44 &  0.086 \\
A3532    & 12~57~21 & -30~21 & 0.0555 & 0 & II-III &   1.44 &  0.085 \\
A3556    & 13~24~07 & -31~40 & 0.0490 & 0 &   I    &   0.48 &  0.060 \\
A3558    & 13~27~57 & -31~29 & 0.0477 & - &   I    &   3.20 &  0.050 \\
A3560    & 13~31~53 & -33~14 & 0.0470 & 3 &   I    &   0.67 &  0.056 \\
A3562    & 13~33~35 & -31~40 & 0.0502 & 2 &   I    &   1.70 &  0.058 \\
A3667    & 20~12~27 & -56~49 & 0.0530 & - &  I-II  &   4.47 &  0.049 \\
A3716    & 20~51~30 & -52~43 & 0.0448 & 1 &  I-II  &   0.52 &  0.037 \\
A3809    & 21~46~59 & -43~53 & 0.0631 & - &   III  &   1.15 &  0.018 \\
A3880    & 22~27~55 & -30~34 & 0.0570 & 0 &   II   &   0.95 &  0.015 \\
A4059    & 23~57~00 & -34~45 & 0.0480 & 1 &   I    &   1.58 &  0.017 \\
IIZW108  & 21~13~56 &  02~33 & 0.0483 & - &    -   &   1.12 &  0.070 \\
MKW3s    & 15~21~52 &  07~42 & 0.0453 & 0 &    -   &   1.37 &  0.035 \\
RXJ0058  & 00~58~55 &  26~57 & 0.0470 & - &    -   &   0.22 &  0.068 \\
RXJ1022  & 10~22~10 &  38~31 & 0.0534 & - &    -   &   0.18 &  0.018 \\
RXJ1740  & 17~40~31 &  35~39 & 0.0430 & - &    -   &   0.26 &  0.026 \\
ZwCl1261 & 07~16~41 &  53~23 & 0.0644 & - &    -   &   0.41 &  0.092 \\
ZwCl2844 & 10~02~36 &  32~42 & 0.0500 & - &    -   &   0.29 &  0.015 \\
ZwCl8338 & 18~10~50 &  49~55 & 0.0473 & - &    -   &   0.40 &  0.043 \\
ZwCl8852 & 23~10~30 &  07~35 & 0.0400 & - &    -   &   0.48 &  0.065 \\
\end{longtable}

\begin{longtable}{lcc|ccccc|ccccc}
\caption{\label{tab:ObsLog} The WINGS-OPT observing log.}\\
\hline\hline
\rm Cluster & Run & Night & \multicolumn{5}{|c|}{$V$ band} & \multicolumn{5}{|c}{$B$ band} \\
 & & & \#exp & $T_{exp}$ & $\mu_{sky}$ & fwhm'' & $m_{lim}$ & \#exp & $T_{exp}$ & $\mu_{sky}$ & fwhm'' & $m_{lim}$ \\
\hline
\endfirsthead
\caption{continued.}\\
\hline\hline
\rm Cluster & Run & Night & \multicolumn{5}{|c|}{$V$ band} & \multicolumn{5}{|c}{$B$ band} \\
 & & & \#exp & $T_{exp}$ & $\mu_{sky}$ & fwhm'' & $m_{lim}$ & \#exp & $T_{exp}$ & $\mu_{sky}$ & fwhm'' & $m_{lim}$ \\
\hline
\endhead
\hline
      A85 &  4N &  15-Sep-01 &  3 & 400 & 21.16 &  1.25 &  24.12 &  3 & 420 & 22.17 &  1.27 &  24.68 \\
     A119 &  4N &  15-Sep-01 &  3 & 400 & 21.32 &  1.20 &  24.29 &  3 & 420 & 22.23 &  1.30 &  24.66 \\
     A133 &  3S &  15-Aug-01 &  3 & 460 & 21.56 &  1.14 &  23.84 &  3 & 420 & 22.36 &  1.38 &  24.06 \\
          &  6S &  12-Jul-02 &  2 & 540 & 21.27 &  1.45 &  22.99 &  2 & 480 & 22.04 &  1.40 &  23.73 \\
     A147 &  1N &  28-Aug-00 &  3 & 400 & 21.36 &  1.15 &  24.34 &  3 & 420 & 22.07 &  1.08 &  24.89 \\
          &  4N &  17-Sep-01 &  3 & 400 & 21.25 &  1.25 &  24.16 &  3 & 420 & 22.22 &  1.24 &  24.75 \\
     A151 &  4N &  16-Sep-01 &  3 & 400 & 21.13 &  1.10 &  24.31 &  3 & 420 & 22.14 &  1.30 &  24.51 \\
     A160 &  4N &  16-Sep-01 &  3 & 400 & 21.23 &  1.10 &  24.42 &  3 & 420 & 22.23 &  1.12 &  24.95 \\
     A168 &  1N &  28-Aug-00 &  3 & 400 & 21.37 &  1.20 &  24.25 &  3 & 420 & 22.21 &  1.16 &  24.80 \\
     A193 &  1N &  28-Aug-00 &  3 & 400 & 21.33 &  1.55 &  23.67 &  3 & 420 & 22.18 &  1.50 &  24.23 \\
          &  4N &  17-Sep-01 &  3 & 400 & 21.29 &  1.18 &  24.36 &  3 & 420 & 22.37 &  1.23 &  24.95 \\
     A311 &  4N &  15-Sep-01 &  3 & 400 & 21.42 &  1.20 &  24.34 &  3 & 420 & 22.32 &  1.23 &  24.82 \\
     A376 &  4N &  16-Sep-01 &  3 & 400 & 21.39 &  1.30 &  24.10 &  3 & 420 & 22.33 &  1.36 &  24.54 \\
     A500 &  5S &  17-Feb-02 &  3 & 460 & 21.48 &  1.00 &  23.98 &  3 & 420 & 22.15 &  1.05 &  24.39 \\
    A548b &  5S &  17-Feb-02 &  3 & 460 & 21.29 &  1.00 &  23.89 &  3 & 420 & 22.13 &  1.05 &  24.38 \\
     A602 &  2N &  26-Apr-01 &  3 & 400 & 20.78 &  1.20 &  24.02 &  3 & 420 & 21.60 &  1.33 &  24.38 \\
     A671 &  2N &  26-Apr-01 &  3 & 400 & 20.90 &  1.30 &  23.90 &  3 & 420 & 21.81 &  1.30 &  24.53 \\
     A754 &  2N &  27-Apr-01 &  3 & 400 & 20.65 &  1.27 &  23.84 &  3 & 420 & 21.43 &  1.25 &  24.43 \\
     A780 &  2N &  29-Apr-01 &  3 & 400 & 20.15 &  1.90 &  22.70 &  3 & 420 & 20.35 &  1.75 &  23.17 \\
          &  6S &  07-Apr-02 &  3 & 460 & 21.20 &  1.21 &  23.50 &  3 & 420 & 22.14 &  1.12 &  24.42 \\
     A957 &  2N &  28-Apr-01 &  3 & 400 & 20.68 &  1.43 &  23.58 &  3 & 420 & 21.21 &  1.73 &  23.62 \\
     A970 &  2N &  28-Apr-01 &  3 & 400 & 20.53 &  1.43 &  23.51 &  3 & 420 & 20.87 &  1.50 &  23.67 \\
          &  6S &  03-Apr-02 &  3 & 460 & 20.15 &  1.24 &  22.93 &  3 & 420 & 20.39 &  1.20 &  23.42 \\
    A1069 &  2N &  27-Apr-01 &  3 & 400 & 20.76 &  1.30 &  23.84 &  3 & 420 & 21.30 &  1.38 &  24.06 \\
    A1291 &  2N &  29-Apr-01 &  3 & 400 & 20.85 &  1.30 &  23.87 &  3 & 420 & 21.47 &  1.34 &  24.30 \\
   A1631a &  5S &  15-Feb-02 &  3 & 460 & 20.99 &  0.85 &  24.14 &  3 & 420 & 21.81 &  1.12 &  24.16 \\
    A1644 &  5S &  15-Feb-02 &  3 & 460 & 21.00 &  0.71 &  24.54 &  3 & 420 & 21.82 &  0.81 &  24.87 \\
    A1668 &  2N &  25-Apr-01 &  3 & 600 & 21.02 &  1.55 &  23.80 &  3 & 600 & 22.04 &  1.40 &  24.70 \\
    A1736 &  5S &  15-Feb-02 &  3 & 460 & 21.18 &  0.74 &  24.53 &  3 & 420 & 21.94 &  0.83 &  24.88 \\
    A1795 &  2N &  26-Apr-01 &  3 & 400 & 21.32 &  1.00 &  24.68 &  3 & 420 & 22.41 &  1.03 &  25.34 \\
    A1831 &  2N &  25-Apr-01 &  3 & 400 & 21.12 &  1.33 &  23.96 &  3 & 600 & 22.13 &  1.20 &  25.08 \\
    A1983 &  2N &  26-Apr-01 &  3 & 400 & 21.38 &  0.93 &  24.87 &  3 & 420 & 22.43 &  0.96 &  25.49 \\
    A1991 &  2N &  26-Apr-01 &  3 & 400 & 21.37 &  1.03 &  24.64 &  3 & 420 & 22.44 &  1.00 &  25.41 \\
    A2107 &  1N &  28-Aug-00 &  3 & 400 & 21.22 &  1.05 &  24.46 &  3 & 420 & 22.16 &  1.23 &  24.66 \\
    A2124 &  2N &  25-Apr-01 &  3 & 400 & 21.08 &  1.03 &  24.50 &  3 & 420 & 22.01 &  1.38 &  24.52 \\
    A2149 &  2N &  27-Apr-01 &  3 & 400 & 19.79 &  1.00 &  23.32 &  3 & 420 & 21.08 &  1.10 &  24.03 \\
    A2169 &  4N &  16-Sep-01 &  3 & 400 & 21.25 &  1.10 &  24.45 &  3 & 420 & 22.36 &  1.80 &  23.98 \\
    A2256 &  2N &  25-Apr-01 &  3 & 400 & 21.12 &  1.21 &  24.17 &  3 & 420 & 22.33 &  1.20 &  24.97 \\
    A2271 &  2N &  27-Apr-01 &  3 & 400 & 19.96 &  1.20 &  23.13 &  3 & 420 & 20.95 &  1.15 &  23.86 \\
    A2382 &  6S &  12-Jun-02 &  2 & 540 & 21.11 &  1.24 &  23.27 &  2 & 480 & 21.85 &  1.40 &  23.64 \\
          &  6S &  13-Jul-02 &  1 & 300 & 21.37 &  1.05 &  23.02 &  1 & 300 & 22.08 &  0.88 &  24.12 \\
    A2399 &  6S &  12-Jun-02 &  2 & 540 & 21.12 &  1.40 &  23.01 &  2 & 480 & 21.87 &  1.52 &  23.47 \\
          &  6S &  13-Jul-02 &  1 & 300 & 21.43 &  1.20 &  22.76 &  1 & 300 & 22.13 &  1.00 &  23.86 \\
    A2415 &  1N &  28-Aug-00 &  3 & 400 & 21.04 &  1.45 &  23.68 &  3 & 420 & 21.99 &  1.23 &  24.57 \\
    A2457 &  4N &  17-Sep-01 &  3 & 400 & 21.11 &  1.35 &  23.92 &  3 & 420 & 22.13 &  1.31 &  24.59 \\
   A2572a &  4N &  17-Sep-01 &  3 & 400 & 21.33 &  1.20 &  24.29 &  3 & 420 & 22.29 &  1.28 &  24.72 \\
    A2589 &  1N &  28-Aug-00 &  3 & 400 & 21.46 &  1.70 &  23.54 &  3 & 420 & 22.29 &  1.65 &  24.08 \\
          &  4N &  17-Sep-01 &  3 & 400 & 21.21 &  0.85 &  24.98 &  3 & 420 & 22.18 &  0.98 &  25.24 \\
    A2593 &  1N &  28-Aug-00 &  3 & 400 & 21.42 &  2.05 &  23.11 &  3 & 420 & 22.22 &  1.60 &  24.11 \\
          &  4N &  17-Sep-01 &  3 & 400 & 21.18 &  0.95 &  24.72 &  3 & 420 & 22.20 &  1.05 &  25.11 \\
    A2622 &  4N &  15-Sep-01 &  3 & 400 & 21.29 &  1.23 &  24.22 &  3 & 420 & 22.28 &  1.26 &  24.75 \\
    A2626 &  1N &  28-Aug-00 &  3 & 400 & 21.49 &  1.64 &  23.63 &  3 & 420 & 22.30 &  1.73 &  23.98 \\
    A2657 &  4N &  16-Sep-01 &  3 & 400 & 21.29 &  1.35 &  24.02 &  3 & 420 & 22.20 &  1.31 &  24.60 \\
    A2665 &  4N &  17-Sep-01 &  3 & 400 & 21.16 &  1.04 &  24.51 &  3 & 420 & 22.12 &  1.16 &  24.85 \\
    A2717 &  3S &  15-Aug-01 &  3 & 460 & 21.69 &  1.33 &  23.57 &  3 & 420 & 22.44 &  1.45 &  23.99 \\
          &  6S &  13-Jun-02 &  2 & 540 & 20.48 &  1.34 &  22.86 &  2 & 480 & 22.26 &  1.28 &  24.14 \\
          &  6S &  13-Jul-02 &  1 & 300 & 21.32 &  0.75 &  23.73 &  1 & 300 & 22.19 &  0.90 &  24.12 \\
    A2734 &  6S &  13-Jun-02 &  2 & 540 & 21.33 &  1.45 &  23.10 &  2 & 480 & 22.14 &  1.22 &  24.19 \\
          &  6S &  14-Jul-02 &  1 & 300 & 21.48 &  1.12 &  22.94 &  1 & 300 & 22.23 &  1.05 &  23.80 \\
    A3128 &  5S &  16-Feb-02 &  3 & 460 & 21.29 &  1.50 &  23.05 &  3 & 420 & 22.00 &  1.33 &  23.88 \\
    A3158 &  5S &  17-Feb-02 &  3 & 460 & 21.44 &  1.38 &  23.31 &  3 & 420 & 21.93 &  1.10 &  24.26 \\
    A3164 &  5S &  15-Feb-02 &  3 & 460 & 21.42 &  1.22 &  23.56 &  4 & 420 & 22.13 &  1.62 &  23.67 \\
    A3266 &  5S &  16-Feb-02 &  3 & 460 & 21.22 &  1.07 &  23.75 &  3 & 420 & 21.99 &  1.17 &  24.15 \\
    A3376 &  5S &  15-Feb-02 &  3 & 460 & 21.14 &  1.14 &  23.53 &  3 & 420 & 21.94 &  1.19 &  24.01 \\
    A3395 &  5S &  17-Feb-02 &  3 & 460 & 21.11 &  0.90 &  24.03 &  3 & 420 & 21.84 &  0.85 &  24.70 \\
    A3490 &  5S &  16-Feb-02 &  3 & 460 & 20.92 &  1.15 &  23.45 &  3 & 420 & 21.73 &  1.07 &  24.22 \\
    A3497 &  5S &  17-Feb-02 &  3 & 460 & 21.39 &  0.69 &  24.79 &  3 & 420 & 22.10 &  0.69 &  25.36 \\
   A3528a &  5S &  15-Feb-02 &  3 & 460 & 20.91 &  0.83 &  24.15 &  3 & 420 & 22.03 &  0.86 &  24.84 \\
   A3528b &  5S &  15-Feb-02 &  1 & 460 & 21.38 &  0.83 &  23.79 &  1 & 180 & 22.09 &  0.92 &  23.63 \\
    A3530 &  5S &  16-Feb-02 &  3 & 460 & 20.99 &  1.06 &  23.66 &  3 & 420 & 21.80 &  1.06 &  24.27 \\
    A3532 &  5S &  16-Feb-02 &  3 & 460 & 21.21 &  0.90 &  24.12 &  3 & 420 & 21.95 &  0.95 &  24.59 \\
    A3556 &  5S &  16-Feb-02 &  3 & 460 & 21.33 &  0.83 &  24.36 &  3 & 420 & 22.08 &  0.83 &  24.94 \\
    A3558 &  5S &  17-Feb-02 &  3 & 460 & 21.35 &  0.86 &  24.29 &  3 & 420 & 21.98 &  0.87 &  24.80 \\
    A3560 &  5S &  17-Feb-02 &  3 & 460 & 21.40 &  0.83 &  24.39 &  3 & 420 & 22.06 &  0.86 &  24.86 \\
    A3562 &  5S &  16-Feb-02 &  1 & 180 & 21.31 &  0.86 &  23.14 &  1 & 180 & 22.01 &  0.95 &  23.53 \\
    A3667 &  6S &  12-Jun-02 &  2 & 540 & 21.39 &  1.60 &  22.84 &  2 & 480 & 22.19 &  1.75 &  23.33 \\
          &  6S &  13-Jul-02 &  1 & 300 & 21.57 &  1.10 &  23.02 &  1 & 300 & 22.28 &  1.00 &  23.93 \\
    A3716 &  6S &  12-Jun-02 &  2 & 540 & 21.33 &  1.65 &  22.75 &  2 & 480 & 22.19 &  1.50 &  23.66 \\
          &  6S &  13-Jul-02 &  1 & 300 & 21.58 &  1.15 &  22.92 &  1 & 300 & 22.32 &  1.15 &  23.64 \\
    A3809 &  6S &  12-Jun-02 &  2 & 540 & 21.24 &  1.50 &  22.91 &  2 & 480 & 22.07 &  1.38 &  23.79 \\
          &  6S &  13-Jul-02 &  1 & 300 & 21.55 &  1.12 &  22.97 &  1 & 300 & 22.30 &  1.10 &  23.73 \\
    A3880 &  3S &  15-Aug-01 &  3 & 460 & 21.63 &  1.45 &  23.35 &  3 & 420 & 22.38 &  1.38 &  24.07 \\
          &  6S &  13-Jun-02 &  2 & 540 & 21.20 &  1.47 &  22.94 &  2 & 480 & 22.00 &  1.21 &  24.04 \\
          &  6S &  13-Jul-02 &  1 & 300 & 21.28 &  1.02 &  23.05 &  1 & 300 & 22.11 &  1.26 &  23.35 \\
    A4059 &  6S &  13-Jun-02 &  2 & 540 & 21.14 &  1.33 &  23.12 &  2 & 480 & 21.97 &  1.25 &  23.95 \\
          &  6S &  13-Jul-02 &  1 & 300 & 21.41 &  0.75 &  23.78 &  1 & 300 & 22.22 &  0.95 &  24.02 \\
  IIZW108 &  4N &  15-Sep-01 &  3 & 400 & 21.23 &  1.60 &  23.62 &  3 & 420 & 22.15 &  1.68 &  24.06 \\
          &  4N &  17-Sep-01 &  4 & 400 & 21.25 &  1.25 &  24.32 &  - &  -  &   -   &   -   &    -   \\
    MKW3S &  2N &  27-Apr-01 &  3 & 400 & 19.87 &  1.11 &  23.26 &  3 & 420 & 21.21 &  1.16 &  24.12 \\
  RXJ0058 &  4N &  16-Sep-01 &  3 & 400 & 21.51 &  1.20 &  24.37 &  3 & 420 & 22.44 &  1.13 &  25.02 \\
  RXJ1022 &  2N &  27-Apr-01 &  3 & 400 & 20.86 &  1.10 &  24.25 &  3 & 420 & 21.84 &  1.16 &  24.79 \\
  RXJ1740 &  2N &  26-Apr-01 &  3 & 400 & 21.48 &  1.10 &  24.55 &  3 & 420 & 22.53 &  1.15 &  25.15 \\
 ZwCl1261 &  4N &  15-Sep-01 &  3 & 400 & 21.14 &  1.08 &  24.43 &  3 & 420 & 22.20 &  1.26 &  24.71 \\
 ZwCl2844 &  2N &  27-Apr-01 &  3 & 400 & 20.77 &  1.03 &  24.35 &  3 & 420 & 21.72 &  1.06 &  24.93 \\
 ZwCl8338 &  2N &  26-Apr-01 &  3 & 400 & 21.52 &  1.03 &  24.72 &  2 & 420 & 22.38 &  1.13 &  24.90 \\
          &  2N &  28-Apr-01 &  1 & 400 & 21.24 &  1.55 &  23.09 &  1 & 420 & 22.25 &  1.45 &  23.92 \\
 ZwCl8852 &  4N &  15-Sep-01 &  3 & 400 & 21.08 &  1.36 &  23.90 &  3 & 420 & 22.02 &  1.43 &  24.35 \\
\end{longtable}

\newpage
\appendix

\section{Basic Data Reduction}
\label{sec:BasDatRed}

The whole reduction procedure has been carried out by means of IRAF-based 
tools. In particular, specially designed IRAF scripts have been
assembled to produce automatically super-bias and super-flat frames for
each observing night \citep{marm1}. The specific tasks related to the treatment of
wide-field imaging (astrometry and mosaicing), even with the
particular layout of the WFC@INT camera, have been managed by the IRAF
mosaic reduction package \verb mscred \citep{vald} and the IRAF
script package \verb wfpred \ developed at the Padova Observatory
(Rizzi and Held, private communication).


The dark current turned out to be always negligible for both the WFC@INT
and WFI@MPG cameras and was not considered in the reduction pipeline.
Similarly, we have not applied fringing corrections, since no
significant fringe patterns are found in both the $B$ and $V$ frames for
either the WFC@INT or WFI@MPG.

\subsection{Bias removal}
\label{sec:BiasCorr}


The bias frames of the WFC@INT camera showed some significant low
frequency structure, with slight systematic differences among
different nights, thus, a 2D bias removal was required.
To produce a reliable and almost noiseless bias frame for
each night (super-bias), we used a specially designed, automatic IRAF
procedure comparing mean, standard deviation and skewness of the
different bias frames and combining only the ones showing 
homogeneous trends. The average scatter of the super-bias counts turned
out to be negligible (0.5--0.8~ADU per pixel: $\sim$7 magnitudes
below the sky surface brightness). We applied the same procedure to
WFI@MPG images, although in this case the bias frames showed more 
constant patterns.

\subsection{Linearity correction}
\label{sec:LinCorr}


Specific tests revealed that the CCDs of the WFC@INT camera
suffer from significant non-linearities over the whole dynamic
range. These have been corrected according to the prescriptions 
given in the CASU INT Wide Field
Survey web-page ({\it http://www.ast.cam.ac.uk/$\sim$wfcsur/foibles.html}).
In order to allow the correction to be performed automatically, the
coefficients of the equations given there have been included in the headers of
WFC images. No linearity problems have been found in the WFI@MPG
detectors.

\subsection{Flat fielding}
\label{sec:FlatCorr}


Dome-flats turned out to be much less stable than sky-flats and were
never used. Again, night super-flats have been produced by an
automatic IRAF script we have devised for this purpose. After bias
subtraction, linearity correction and trimming of the flats, this
procedure rejected the low--counts and close--to--saturation flats;
then a single, normalized super-flat was produced for each filter,
combining those flats whose marginal distributions of counts along
both the X and the Y axes had similar values of mean, standard
deviation and skewness. Both the random (pixel by pixel) variance and
the systematic differences among flats taken on the same nights turned out
to be less than 1\% (0.01 mag). Sky flats taken on different nights of
the same observing run usually showed a good mutual agreement, while 
significant differences have been found in the patterns
of flats taken in different runs.

Due to non-uniform illumination, the WFI@MPG camera has been reported
to show significant large-scale spatial gradients in photometry across
the entire field of view, and across each of its eight chips
individually \citep{manf,koch}. This problem
cannot be solved by usual flat fielding because the illumination
unevenness affects both flats and science
exposures alike. 
In Section~\ref{sec:PhotCal} this problem is faced and solved by means
of a 2nd-order, 2D polynomial fit of the photometric residuals over the fields.

\subsection{Astrometry}
\label{sec:AstroCorr}

\begin{figure*}
\vskip -5truecm
\begin{center}
\vskip 1truecm
\includegraphics[width=19cm]{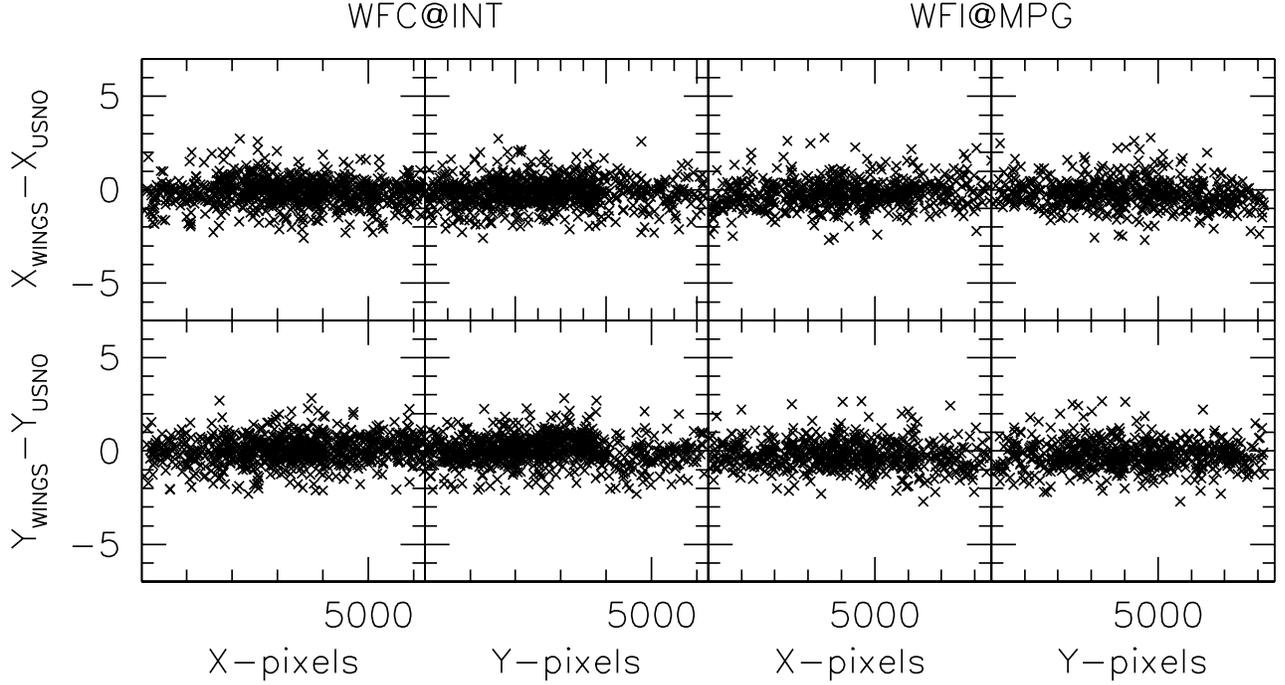}
\vskip -5truecm
\caption{X-pix and Y-pix differences between WINGS and USNO coordinates
of galaxies, after having applied our astrometric solutions for WFC@INT 
(left) and WFI@MPG (right).}
\label{fig:Astrom}
\end{center}
\end{figure*}

Finding an astrometric solution adequate for the proposed scientific
objectives is a specific and critical task to be addressed when
dealing with wide--field imaging. Usually, the wider the field, the
larger the geometric distortions introduced by the optical layout of
the camera. It is important to note that, besides the astrometric
measurements, such distortions can also significantly affect the
photometry, due to the mis-shaped smearing of the light on the
pixel array. In order to map, model and correct distortions in
wide--field images, one has to compare physical (pixels) and world
($\alpha$, $\delta$) coordinates for a given sample of point-like
sources (stars) in the field. Strong distortions require sizeable
astrometric samples of stars uniformly spread throughout the field. 
Since such samples are seldom available, it is often convenient to
adopt an astrometric solution obtained once and for all from a suitable
astrometric field containing several hundred (or even thousands of)
stars. 

The WFC@INT imaging is well known to be affected by strong geometric
distortions. The astrometric solutions for the two filters $B$ and $V$
have been obtained using the astrometric regions ACR-D and ACR-N
\citep{ston}. These solutions have been applied (after re-centering) 
to each northern cluster and standard field.

For the WFI@MPG camera, a precise astrometric solution,
obtained using the astrometric regions ACR-E and ACR-M \citep{ston}, 
was already available (Rizzi and Held, private communication). In this
case, only the re-centering step was performed.

Figure~\ref{fig:Astrom} shows the differences (in pixel units) between
WINGS and USNO \citep{mone} coordinates of galaxies for three
clusters observed with WFC@INT (A85, A119, A168; left panels) and
three more clusters observed with WFI@MPG (A500, A3395, A3490; right
panels). The comparison is performed using galaxies since the stars of
the USNO database are usually saturated in our imaging. Since
centering algorithms are likely to be much less precise for galaxies than
for stars, the formal precision obtained from the astrometric solution
applied to the stars of the astrometric fields ($r.m.s.\sim$0.1~pix)
turn out to be much smaller than that found for galaxies and shown
in Figure~\ref{fig:Astrom} ($r.m.s.\sim$0.75~pix, corresponding to 0.25
and 0.18 arcseconds in the case of WFC@INT and WFI@MPG, respectively).
Still, this is accurate enough to ensure a precise pointing for
the multi-fiber spectroscopy carried out in the framework of the
WINGS-SPE survey with both WYFFOS@WHT (fiber
of 1.6 arcseconds) and 2dF@AAT (fiber of 2.1 arcseconds).

\section{Photometric Calibration}
\label{sec:PhotCalProc}

\subsection{Formalism}
\label{sec:PhotMethod}

\begin{figure*}
\vskip -1.5truecm
\begin{center}
\hskip -1truecm
\vskip 1truecm
\includegraphics[width=18cm]{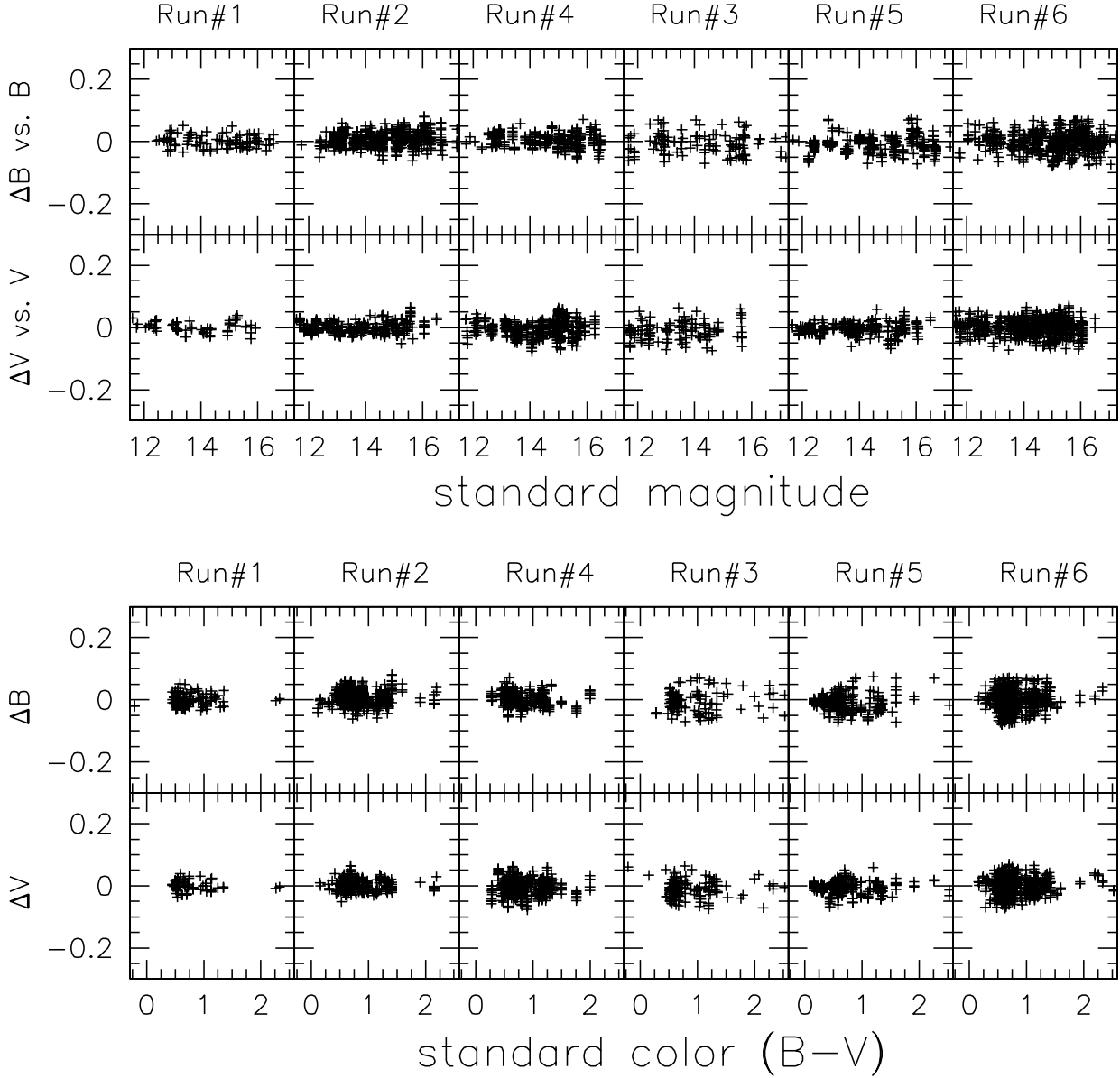}
\vskip -1truecm
\caption{Residuals of our photometric
calibration for each observing run and for each band, as a
function of both standard magnitudes and colors.}
\label{fig:Calib}
\end{center}
\end{figure*}

Following \citet[ see also \citealt{vare}]{mole}, we assume
that, even though the atmospheric extinction varies night by night,
the observing set remains stable during each observing run. This
implies that the out-of-atmosphere instrumental magnitude of each
standard star measured on a given chip of the mosaic, is constant
throughout the run, being different on different chips. Thus, for each
observing run and for each filter, this procedure provides us with an
extinction coefficient for each night and with a set of zero points
and color coefficients (one pair for each CCD of the mosaic) holding
for the entire run.

First, we looked for the extinction coefficients, solving the following
system of N generalized Bouger's equations:

\begin{equation}
m^\mu_{scn}= m_{0;sc}+k_nX^\mu\ \ \ ; \ \ \mu = 1,...,N
\label{eq:Ext}
\end{equation}

\noindent
where $k$ is the extinction coefficient, $X$ is the airmass and $N$ is
the number of determinations ($m$) obtained for the out-of-atmosphere
instrumental magnitude ($m_0$) of a given star ($s$) during a given
night ($n$) on a given chip of the mosaic ($c$).

For the sake of formal simplicity, in the minimization algorithm
the equations~(\ref{eq:Ext}) have been expressed in the form:

\begin{equation}
m^\mu_{scn}= \sum_{p,q}{m_{0;pq}\delta_{ps}\delta_{qc}}+\sum_r{k_rX^\mu_r\delta_{rn}},
\label{eq:Lmf}
\end{equation}

\begin{table}[tb]
  \centering
\caption{Extinction coefficients of the WINGS-OPT survey. The
  extinction coefficients of the Run~\#6 were fixed to
  the values given in the ESO web site because the scarcity of
  measurements made it impossible computing this values from the
  standard observations.}
\label{tab:Calib1}
  \begin{tabular}{cccc}
\hline\hline
 Run &  Night & $k^V$ &$ k^B$ \\
\hline
 \#1  &   28-Aug-2000  & 0.115 & 0.198 \\
\hline
 \#2  &   25-Apr-2001  & 0.117 & 0.225 \\ 
      &   26-Apr-2001  & 0.113 & 0.249 \\ 
      &   27-Apr-2001  & 0.106 & 0.259 \\ 
      &   28-Apr-2001  & 0.123 & 0.246 \\ 
      &   29-Apr-2001  & 0.122 & 0.255 \\ 
\hline
 \#3  &   15-Aug-2001  & 0.116 & 0.179 \\              
      &   16-Aug-2001  & 0.118 & 0.188 \\              
\hline
 \#4  &   15-Sep-2001  & 0.099 & 0.195 \\
      &   16-Sep-2001  & 0.133 & 0.252 \\
      &   17-Sep-2001  & 0.109 & 0.199 \\
\hline
 \#5  &   14-Feb-2002  & 0.091 & 0.158 \\
      &   15-Feb-2002  & 0.092 & 0.159 \\
      &   16-Feb-2002  & 0.086 & 0.147 \\
\hline
 \#6  &   Apr-Jul-2002 & 0.150 & 0.200 \\
      &   Aug-2002     & 0.092 & 0.238 \\
\hline
\end{tabular}
\end{table}

\noindent
where the indices \{$p,q,r$\} respectively span all possible values of
\{$s,c,n$\} and $\delta$ is the Kronecker symbol.
Table~\ref{tab:Calib1} shows the extinction coefficients obtained in
this way for each filter in each observing night (see the table caption
as far as the Run~\#6 is concerned).
Table~\ref{tab:Calib2} reports, for each observing run and for each
filter, the photometric zero points $Z_c$ and the color coefficients
$C_c$ of the different mosaic CCDs($c$). 
Each pair of coefficients is
obtained solving a system of equations like this:

\begin{equation}
(m_{0;sc}-m^{std}_s)=Z_c+C_c\times (B-V)^{std}_s,
\label{eq:Stand}
\end{equation}

\noindent
where the out-of-atmosphere instrumental magnitudes $m_{0;sc}$ are
known from equations~\ref{eq:Lmf}, $m^{std}$ and $(B-V)^{std}$ are
the standard magnitudes and colors taken from the \citet{land}
catalogs and $c$ and $s$ span all possible chips and stars,
respectively. 

Figure~\ref{fig:Calib} illustrates the results of our calibration procedure
applied to the WINGS-OPT standard stars in both the B and V bands. In particular,
the residuals (eq.~\ref{eq:Stand}) are reported as a function of both 
standard magnitudes and colors. 

During run~\#6-Apr. (service mode), no standard stars were observed
in the CCD~\#8. We used for this run the $Z_c$ and $C_c$ coefficients
of run~\#6-June (also reported in Table~\ref{tab:Calib2}). It is 
worth noting, however, that the observations from run~\#6-Apr. (A780 
and A970) have been used to just compare the photometry between WFC@INT 
and WFI@ESO (see Figure~\ref{fig:CompPhot1}).

\subsection{Photometric normalization of the Mosaics}
\label{sec:MosaicProc}

In order to properly run SExtractor \citep{bert} over the
co-added mosaic frames, we processed them as follows:

First, each CCD of each multi--extension image has been divided by the exposure time
and diminished by the mode of the histogram of the pixel counts,
assumed to be a rough estimate of the average sky value. In
Section~\ref{sec:BackRem}, the final, much more accurate procedure we used
for backgroung subtraction is outlined. Provisionally, the mode subtraction
provided a flat, close to zero background over the whole
image, allowing us to perform the next step of the normalization
procedure, that is the correction for both atmospheric extinction and
gain differences among the different CCDs.


For each observing run and for each filter, this is obtained by
multiplying each pixel of the mosaic image by the factor:
\begin{equation}
10^{-0.4 [(Z_c-<Z>)-k_n (X-1)]}
\label{eq:Zcorr}
\end{equation}
where $n$ and $c$ are the night and the CCD which the pixel comes
from, while $<$Z$>$ and $k_n$ represent the CCD-averaged
zero point of the run and the extinction coefficient of the
night, respectively. The newly derived image can be processed as
a whole by SExtractor, using the virtual zero point:
\begin{equation}
Z_{SEx}=<Z>-k_n.
\label{eq:ZSEx}
\end{equation}
Since the color terms of the photometric calibration are not considered in 
the previous procedure, the magnitudes derived in this way by SExtractor ($m_{SEx}$)
have to be color-corrected afterwards, depending on the true colors of the
object, as well as on the CCD($c$) where it is located:
\begin{equation}
m=m_{SEx}+C_c\times (B-V).
\label{eq:ColCorr}
\end{equation}
In this formula $C_c$ represents the color coefficient of the
particular CCD and the true color $(B-V)$ can be easily evaluated by:
\begin{equation}
(B-V)=(B_{SEx}-V_{SEx})/(1-\delta C_c),
\label{eq:ColCorr1}
\end{equation}
where $\delta C_c=C^B_c-C^V_c$.


The headers of the co-added mosaic frames of each cluster have been
updated with keywords giving the proper photometric coefficients
(including $Z_{SEx}$ and $\delta C_c$) and the subtracted background
values.

It is worth noting that the mosaic frames obtained by co-addition of
exposures taken on different observing nights (some clusters of
run~\#6), possibly with different calibration coefficients $Z_c$ and
$C_c$, in principle cannot be processed as explained above. The
procedure we used in this case, that is to adopt weight-averaged calibration
coefficients, is likely to be only a crude approximation. Therefore,
even though these mosaics can be useful for surface photometry and
morphology, they cannot be trusted as far as the absolute photometry
is concerned.

\begin{longtable}{lc|cc|cc}
\caption{\label{tab:Calib2} Calibration coefficients of the WINGS-OPT survey.}\\
\hline\hline
   Run  &  Chip($c$)  & \multicolumn{2}{|c|}{$V$ band} & \multicolumn{2}{|c}{$B$ band} \\
        &             & $Z_c$ & $C_c$ & $Z_c$ & $C_c$ \\
\hline
\endfirsthead
\caption{continued.}\\
\hline\hline
   Run  &  Chip($c$)  & \multicolumn{2}{|c|}{$V$ band} & \multicolumn{2}{|c}{$B$ band} \\
        &             & $Z_c$ & $C_c$ & $Z_c$ & $C_c$ \\
\hline
\endhead
\hline
    \#1       &  1 &   24.61  &  -0.023  &    24.63  &  0.039 \\
              &  2 &   24.61  &  -0.022  &    24.62  &  0.034 \\
              &  3 &   24.57  &  -0.013  &    24.57  &  0.071 \\
              &  4 &   24.60  &  -0.011  &    24.59  &  0.029 \\
\hline
     \#2      &  1 &   24.74  &  -0.001  &    25.00  &  0.093 \\
              &  2 &   24.71  &   0.001  &    25.00  &  0.122 \\
              &  3 &   24.70  &   0.016  &    24.98  &  0.136 \\
              &  4 &   24.71  &   0.001  &    24.98  &  0.114 \\
\hline
     \#3      &  1 &   24.18  &  -0.072  &    24.73  &  0.162 \\
              &  2 &   24.24  &  -0.114  &    24.72  &  0.192 \\
              &  3 &   24.24  &  -0.149  &    24.70  &  0.186 \\
              &  4 &   24.24  &  -0.059  &    24.72  &  0.202 \\
              &  5 &   24.20  &  -0.073  &    24.69  &  0.259 \\
              &  6 &   24.16  &  -0.066  &    24.65  &  0.251 \\
              &  7 &   24.20  &  -0.092  &    24.66  &  0.214 \\
              &  8 &   24.23  &  -0.056  &    24.71  &  0.221 \\
\hline
     \#4      &  1 &   24.76  &  -0.063  &    24.77 &   0.044 \\
              &  2 &   24.70  &   0.020  &    24.78 &   0.031 \\
              &  3 &   24.68  &   0.015  &    24.75 &   0.081 \\
              &  4 &   24.71  &  -0.002  &    24.80 &   0.026 \\
\hline
     \#5      &  1 &    24.08 &  -0.079 &   24.51  &   0.278  \\
              &  2 &    24.04 &  -0.069 &   24.51  &   0.239  \\
              &  3 &    24.04 &  -0.067 &   24.46  &   0.311  \\
              &  4 &    24.10 &  -0.078 &   24.52  &   0.282  \\
              &  5 &    24.07 &  -0.038 &   24.50  &   0.273  \\
              &  6 &    24.04 &  -0.085 &   24.48  &   0.249  \\
              &  7 &    24.03 &  -0.082 &   24.47  &   0.259  \\
              &  8 &    24.12 &  -0.103 &   24.58  &   0.247  \\
\hline		   			                      
 \#6-Apr      &  1 &    24.19 &  -0.080 &   24.79  &    0.200 \\
              &  2 &    24.18 &  -0.114 &   24.73  &    0.212 \\
              &  3 &    24.17 &  -0.102 &   24.74  &    0.213 \\
              &  4 &    24.21 &  -0.083 &   24.78  &    0.230 \\
              &  5 &    24.17 &  -0.043 &   24.75  &    0.243 \\
              &  6 &    24.13 &  -0.072 &   24.71  &    0.220 \\
              &  7 &    24.12 &  -0.086 &   24.70  &    0.220 \\
              &  8 &    24.13 &  -0.036 &   24.67  &    0.294 \\   
\hline		   			                      
 \#6-Jun      &  1 &    24.19 &  -0.083 &   24.80  &    0.183 \\
              &  2 &    24.09 &  +0.032 &   24.73  &    0.220 \\
              &  3 &    24.21 &  -0.133 &   24.77  &    0.201 \\
              &  4 &    24.17 &  -0.037 &   24.76  &    0.243 \\
              &  5 &    24.22 &  -0.138 &   24.77  &    0.200 \\
              &  6 &    24.07 &  -0.023 &   24.69  &    0.215 \\
              &  7 &    24.10 &  -0.049 &   24.67  &    0.268 \\
              &  8 &    24.13 &  -0.036 &   24.67  &    0.294 \\
\hline		   			                      
 \#6-Jul      &  1 &    24.16 &  -0.077 &   24.71  &    0.229 \\
              &  2 &    24.10 &  -0.008 &   24.73  &    0.223 \\
              &  3 &    24.09 &  -0.048 &   24.74  &    0.179 \\
              &  4 &    24.15 &  -0.031 &   24.75  &    0.239 \\
              &  5 &    24.16 &  -0.083 &   24.76  &    0.180 \\
              &  6 &    24.08 &  -0.045 &   24.69  &    0.251 \\
              &  7 &    24.09 &  -0.059 &   24.65  &    0.286 \\
              &  8 &    24.11 &  -0.010 &   24.68  &    0.295 \\
\hline		   			                      
 \#6-Aug      &  1 &    24.17 &  -0.176 &   24.87  &    0.090 \\
              &  2 &    24.01 &  -0.041 &   24.70  &    0.271 \\
              &  3 &    24.01 &  -0.054 &   24.70  &    0.242 \\
              &  4 &    24.00 &  +0.025 &   24.69  &    0.341 \\
              &  5 &    24.06 &  -0.090 &   24.72  &    0.214 \\
              &  6 &    24.00 &  -0.081 &   24.67  &    0.231 \\
              &  7 &    24.00 &  -0.078 &   24.55  &    0.370 \\
              &  8 &    24.00 &  -0.002 &   24.59  &    0.334 \\
\end{longtable}

\end{document}